\pdfoutput=1
\documentclass[twocolumn,showpacs,preprintnumbers,final,aps,citeautoscript,footinbib,prb,superscriptaddress,longbibliography,
]{revtex4-2}
\usepackage{amsmath,amssymb,amsfonts}
\usepackage{physics}
\usepackage{graphicx}
\graphicspath{{./figures/}}
\usepackage{float}
\usepackage{bm}
\usepackage{times}
\usepackage[colorlinks,bookmarks=true,citecolor=blue,linkcolor=blue,urlcolor=blue,breaklinks=true]{hyperref}
\usepackage{amsthm}
\usepackage{mathrsfs}
\usepackage{mleftright}
\usepackage{mathtools}
\usepackage{comment}

\newcommand{\rmm}[1]{{\rm{#1}}}
\newcommand{\bmm}[1]{{\bm{#1}}}

\newcommand{\be}{\mathrm{e}}

\newcommand{\lr}[3] { \left#1 #2 \right#3}
\newcommand{\hc}[1]{ {#1} ^{\dagger} }

\newcommand{\com}[2]{\lr{[}{#1,#2}{]} }

\newcommand{\chuukakko}[1]{\lr{\{}{#1}{\}}}

\newcommand{\acom}[2]{\chuukakko{#1,#2}}

\newcommand{\hf}{\frac{1}{2}}

\newcommand{\eps}{\epsilon}

\newcommand{\sx}{\sigma^{x}}
\newcommand{\sy}{\sigma^{y}}
\newcommand{\sz}{\sigma^{z}}

\newcommand{\im}{\rmm{i}}
\newcommand{\nt}{\notag \\}

\newcommand{\FP}{\qty(-1)^F}

\begin{document}
\title{Emergent spacetime supersymmetry in an interacting Kitaev chain with explicit supersymmetry}
\preprint{Phys. Rev. B \textbf{111}, 075136 (2025); DOI: 10.1103/PhysRevB.111.075136}
\author{Urei Miura} 
\email{urei.miura@yukawa.kyoto-u.ac.jp}
\affiliation{%
Division of Physics and Astronomy, Graduate School of Science, 
Kyoto University, Kyoto 606-8502, Japan}%
\affiliation{Center for Gravitational Physics and Quantum Information, Yukawa Institute for Theoretical Physics, Kyoto University, Kitashirakawa Oiwake-Cho, Kyoto 606-8502, Japan}%
\author{Keisuke Totsuka}
\affiliation{Center for Gravitational Physics and Quantum Information, Yukawa Institute for Theoretical Physics, Kyoto University, Kitashirakawa Oiwake-Cho, Kyoto 606-8502, Japan}%
\date{Received 25 October 2024; accepted 5 February 2025; published 18 February 2025}
\begin{abstract}
We investigate the emergence of spacetime supersymmetry (SUSY) in an interacting Kitaev chain model with explicit microscopic $\mathcal{N}=1$ quantum mechanical SUSY. As the interaction strength is varied, the model transitions from a weak-coupling gapless phase with spontaneously broken SUSY to a strong-coupling phase with restored SUSY.
In this paper, we numerically determine the transition point and investigate the phase structure around it. The weak-coupling phase is governed by the Ising conformal field theory (CFT) with central charge $c=1/2$, which agrees with the prediction made in our previous paper. At the SUSY restoration transition, which borders the strong-coupling gapped phase, the transition belongs to the $c=7/10$ tricritical Ising universality with emergent superconformal invariance.
Crucially, the phase structure around the transition closely aligns with the scenario proposed by Zamolodchikov and others based on the integrable deformation of CFTs.
These findings provide a concrete example of a lattice model where explicit microscopic supersymmetry at criticality enriches the understanding of phase transitions governed by supersymmetric conformal field theories.
\end{abstract}
\maketitle
\section{INTRODUCTION}
Supersymmetry (SUSY) \cite{WESS197439,witten1981dynamical,witten1982constraints} has long been a subject of theoretical interest, not only in the realm of particle physics \cite{weinberg1976implications,PhysRevD.14.1667} but also within condensed matter physics. Classic examples include the application of SUSY to handle systems with quenched disorder \cite{parisi1979random,Efetov-book-96} and strong correlation \cite{PhysRevLett.60.821,PhysRevLett.64.2567}, as well as the emergence of SUSY in two-dimensional critical phenomena \cite{Friedan-Q-S-SUSY-85,Qiu-86}. 
More recently, SUSY has been discussed in the contexts of ultracold atoms \cite{yu2008supersymmetry,yu2010simulating} and generalized Josephson junctions \cite{PhysRevB.90.075408,PhysRevLett.123.026401}. 
SUSY is often explored using Bose-Fermi mixtures where both bosonic and fermionic degrees of freedom are incorporated explicitly as atoms with different statistics. Notably, experiments that could detect the Nambu-Goldstone (NG) fermions (Goldstinos) have been proposed theoretically \cite{PhysRevB.110.064512} (for other related results, see the references cited therein).

However, the coexistence of fermionic and bosonic particles is not a prerequisite for SUSY. Even purely fermionic systems can be supersymmetric \cite{nicolai1976supersymmetry,nicolai1977extensions,PhysRevLett.90.120402,fendley2003lattice}. 
In particular, Nicolai's pioneering study \cite{nicolai1976supersymmetry,nicolai1977extensions} and the lattice realization of $\mathcal{N}=2$ superconformal field theories (SCFTs) \cite{PhysRevLett.90.120402,fendley2003lattice} are seminal examples of supersymmetry realized in purely fermionic systems. These models, along with subsequent developments such as the ``hard-core fermion'' models derived from them, exhibit a remarkable phenomenon known as superfrustration, which refers to extensive ground-state degeneracy, irrespective of the system’s dimensionality or boundary conditions \cite{fendley2005exact,PhysRevLett.101.146406,huijse2008superfrustration,huijse2012supersymmetric,fendley2019free,Chepiga-M-S-21}. Moreover, extensions of Nicolai's models have been proposed to study spontaneous supersymmetry breaking, revealing SUSY SSB phase transitions and the appearance of NG fermions \cite{sannomiya2016supersymmetry,sannomiya2017supersymmetry,moriya2018ergodicity,moriya2018supersymmetry,sannomiya2019supersymmetry,sannomiyaDron,Katsura_2020,SUSYfracton,miura2023supersymmetry,miura2024interacting}.
Additionally, supersymmetry can also manifest in topological superconductors and systems composed of Majorana fermions with translational symmetry, whereas the supercharge is implemented in a nonlocal way \cite{grover2014emergent,hsieh2016all}. The physical implications of these systems and their potential applications have also been explored \cite{marra20221d,PhysRevB.105.214525,marra2024majorana}.
Including such nonlocal examples, supersymmetry occurs widely in condensed matter physics, making it essential to study its phase structure, quantum phase transitions, and low-energy excitations.

Nevertheless, a general understanding of low-energy excitations in lattice supersymmetric systems remains less developed compared to their bosonic counterparts \cite{PhysRevLett.110.091601,PhysRevLett.108.251602}. Existing studies on fermionic systems, such as the emergence of supersymmetry at low energies in the Majorana-Hubbard model \cite{rahmani2015emergent,rahmani2019interacting,PhysRevB.92.235123}, and pseudo-supersymmetric models like the one discussed by O’Brien and Fendley \cite{o2018lattice}, have focused on either emergent or approximate supersymmetry. 

In contrast, in Ref.~\onlinecite{miura2024interacting}, a model of lattice interacting Majorana fermions with explicit $\mathcal{N}=1$ supersymmetry has been introduced, providing an opportunity to explore the phases of fermionic systems governed by supersymmetry. The model is fundamentally different from previous Majorana-Hubbard \cite{rahmani2015emergent,rahmani2019interacting,PhysRevB.92.235123} and O'Brien-Fendley models \cite{o2018lattice} in that it possesses exact $\mathcal{N}=1$ supersymmetry at the microscopic level, ensuring that both the Hamiltonian and the system's defining operators satisfy SUSY relations \cite{witten1981dynamical,witten1982constraints} at all energy scales. This allows for precise exploration of phenomena such as spontaneous supersymmetry breaking (SUSY SSB) \cite{witten1981dynamical,witten1982constraints} and the associated phase transitions. 

As the interaction strength increases, this model interpolates between the critical Kitaev chain \cite{kitaev2001unpaired} of non-interacting gapless Majorana fermions and the Fendley model of interacting gapless Majorana fermions \cite{fendley2019free}. The variational and mean-field analyses, as well as the exact ground states at the frustration-free (F.F.) point, suggest that weak interactions lead to spontaneous supersymmetry breaking with the appearance of a NG fermion, whereas supersymmetry is restored in the strong coupling regime. This NG fermion corresponds to the gapless excitations in the critical Kitaev chain, as predicted by mean-field approximations. 
While the phase structure of this model has been studied \cite{miura2024interacting}, the precise determination of the critical point for SUSY restoration and its universality class remains an open question, with only a variational bound and qualitative mean-field description below the transition available. Additionally, the nature of the phase after SUSY restoration is largely unknown, except at the F.F. point, where the ground states are explicitly known.

\begin{figure}[htb]
  \includegraphics[width=\columnwidth,clip]{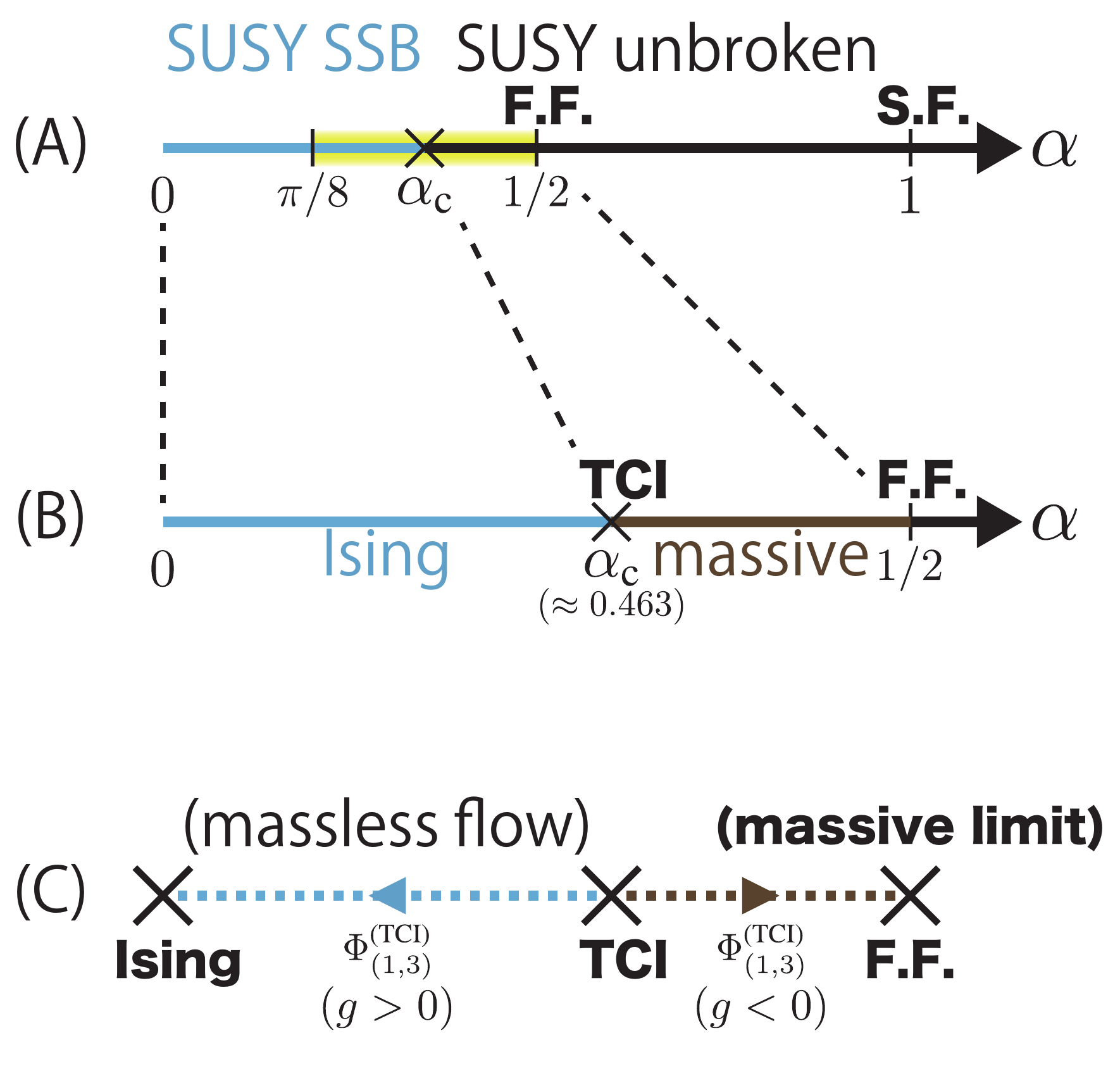}
  \caption{
(a) The schematic phase diagram of the model \eqref{eq:Ham} proposed in Ref.~\onlinecite{miura2024interacting}, (b) the refined one (for $0 \leq \alpha \leq 1/2$) obtained 
in this paper, and (c) the corresponding renormalization group (RG) interpretation. The ``F.F.'' stands for a frustration-free point at which the ground states are explicitly known \cite{miura2024interacting}, and (at least) at $\alpha =1$, superfrustration (``S.F.'') with extensive ground-state degeneracy occurs \cite{fendley2019free}. 
When $0 \leq \alpha < \alpha_{\text{c}}$, supersymmetry is broken spontaneously (``SUSY SSB'') and the resulting gapless fermion (the Nambu-Goldstone fermion or Goldstino) dominates low-energy physics, while at least for $\alpha_{\text{c}} \leq \alpha \leq 1/2$, SUSY is unbroken (``SUSY unbroken''). 
The SUSY restoration transition point $\alpha =\alpha_{\text{c}}$ which has been shown to exist in a window 
$\pi/8 \leq \alpha_{\text{c}} < 1/2$ \cite{sannomiyaDron,miura2024interacting} is numerically determined as $\alpha_{\text{c}} \approx 0.463$. 
The transition which belongs to the universality of the two-dimensional tricritical Ising (TCI) model (with $c=7/10$) is driven by a single relevant operator; depending on the sign of the coupling, the system flows either to the $c=1/2$ critical Ising phase with broken SUSY or to a two-fold degenerate massive phase with unbroken SUSY, which is adiabatically connected to the F.F. point $\alpha=1/2$ [(c); see Sec.~\ref{sec:perturbed-CFT} for the detail]. 
\label{Fig-PhaseDiagram}}
\end{figure}

In this study, we numerically investigate these problems, paying particular attention to the phase structure around the transition point. 
The main results are summarized in Fig.~\ref{Fig-PhaseDiagram} (B). The analyses of the ground-state energy density, the excitation gap, the central charge, and the universal gap ratio \cite{rahmani2015emergent,rahmani2019interacting,PhysRevB.92.235123} all suggest that the SUSY restoration transition, which belongs to the universality class of the two-dimensional tricritical Ising (TCI) model with a central charge $c=7/10$ \cite{Friedan-Q-S-SUSY-85,Qiu-86}, occurs at the interaction strength $\alpha=\alpha_{\text{c}} \approx 0.463$. 
A single symmetry-allowed relevant operator (the vacancy chemical potential or the $t$-field) drives the system either to the extended critical Ising phase ($c=1/2$) with broken SUSY ($\alpha<\alpha_{\text{c}}$) or to a two-fold degenerate massive phase with restored SUSY ($\alpha>\alpha_{\text{c}}$), thereby providing a lattice realization of the scenario proposed in Refs.~\onlinecite{Zamolodchikov-c-87,Kastor-M-S-89}.

The structure of this paper is as follows:
In Sec.~\ref{sec:model}, we introduce the model proposed in our previous paper \cite{miura2024interacting} and provide the theoretical background necessary to interpret the numerical results. We define the supercharge and Hamiltonian of the system and discuss the conditions under which SUSY is either broken or preserved.
In Sec.~\ref{sec:critical-pt-by-numerics}, we numerically determine the critical point for SUSY restoration using finite-size DMRG and exact diagonalization. We analyze the ground-state energy density and energy gap to identify the transition point.
In Sec.~\ref{sec:behavior_at_transition_point}, we investigate the critical behavior at the phase transition. We estimate the central charge from entanglement entropy and calculate universal gap ratios, confirming the system's universality and the role of TCI CFT at the critical point.
In Sec.~\ref{CONCLUSION}, we summarize our findings.
In Appendix~\ref{Spin Representation of the Hamiltonian via Jordan-Wigner Transformation}, we provide the spin representation of the Hamiltonian through the Jordan-Wigner transformation.
In Appendix~\ref{low_energy_behavior_in_fermion_TCI}, we discuss the low-energy behavior of the fermion theory at the TCI and Ising critical points \cite{Zamolodchikov-TCIM-Ising-91}.

\section{MODEL and Theoretical Background}
\label{sec:model}
In this section, we introduce the model proposed in our previous paper \cite{miura2024interacting} and outline the theoretical background necessary for interpreting the numerical results. For further details, refer to \cite{miura2024interacting}. Unless otherwise specified, we assume periodic boundary conditions in the fermion representation.

\subsection{Supercharge and Hamiltonian}
\label{subsec:Supercharge and Hamiltonian}
The $\mathcal{N}=1$ quantum mechanical supersymmetry algebra is defined through the following commutation relations among three Hermitian operators, $R$, $\FP$, and $H$ \cite{witten1981dynamical}:
\begin{align}
\label{eq:1}
    &H=\hf R^2,\nt
    &\acom{R}{\FP}=0,\nt
    &\qty(\FP)^2=+1.
\end{align}
Here, $R$ represents the \textit{real} supercharge, $\FP$ denotes the fermion parity, and $H$ is the system's Hamiltonian \cite{witten1981dynamical}. In systems with $\mathcal{N}=1$ quantum mechanical supersymmetry, the Hamiltonian is determined by the supercharge. From the Eq.~\eqref{eq:1}, we have $\com{H}{R}=0$, implying that the system is supersymmetric, and $\com{H}{\FP}=0$, allowing the energy eigenstates to also be eigenstates of the fermion parity. For more details on the energy spectrum of supersymmetric systems, see Ref.~\cite{witten1981dynamical}.

\begin{figure}[H]
  \includegraphics[width=\columnwidth,clip]{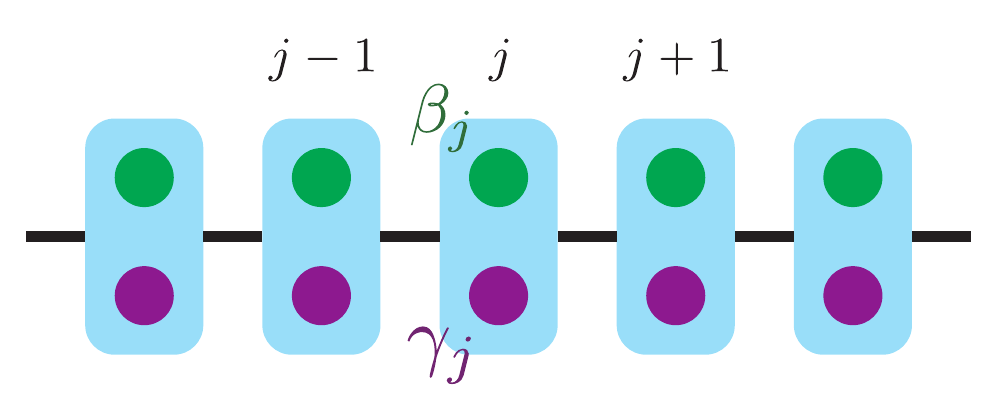}
  \caption{Schematic of the system. Each site hosts Majorana fermions $\beta_{j},\gamma_{j}$, forming a qubit. 
  \label{fig:02}}
\end{figure}

We consider a one-dimensional periodic lattice with $L$ sites, defining Majorana fermions $\beta_j\qty(=\hc{\beta}_j), \gamma_j\qty(=\hc{\gamma}_j)$ as follows:
\begin{align}
    &\acom{\beta_{i}}{\beta_{j}}=\acom{\gamma_{i}}{\gamma_{j}}=2\delta_{i,j}, \nt
    &\acom{\beta_{i}}{\gamma_{j}}=0.  
\end{align}
This setup defines a fermion Fock space (one qubit per site), as depicted in Fig.~\ref{fig:02}.
Next, we define the real supercharge $R\qty(\alpha)$ and fermion parity $\FP$ as follows:
\begin{equation}
\begin{split}
& R\qty(\alpha)\coloneqq\sum_{j=1}^{L} \qty{\qty(1-\alpha)\beta_j+\alpha \im \beta_j\beta_{j+1}\gamma_j}
\quad (0\leq \alpha\leq1),  \\
& \FP\coloneqq\prod_{j=1}^L \im \beta_j \gamma_j  \; .
\end{split}
\label{eq:2}
\end{equation}
Here, $\beta_{L+j}\coloneqq\beta_{j}$ and $\gamma_{L+j}\coloneqq\gamma_{j}$ for periodic boundary conditions. The Hamiltonian $H\qty(\alpha)$ is computed as the square of the supercharge \eqref{eq:2}:
\begin{equation}
\begin{split}
H\qty(\alpha) & \coloneqq\hf R\qty(\alpha)^2 \\ 
& =  \qty(1-\alpha)^2 H_0 + \alpha\qty(1-\alpha) H_{\text{CKC}} + \alpha^2 H_{\text{Fen}} \; .
\end{split}
\label{eq:Ham}
\end{equation}
Here,
\begin{align}
    &\label{eq:3}
    H_0\coloneqq\frac{L}{2},\\
    &\label{eq:4}
    H_{\text{CKC}}\coloneqq\sum_{j=1}^{L} \qty(\im \beta_{j+1}\gamma_j-\im\beta_{j-1}\gamma_{j-1}),\\
  \begin{split}  
    &\label{eq:5}
    H_{\text{Fen}}\coloneqq\hf\qty(\sum_{j=1}^{L}\im \beta_j\beta_{j+1}\gamma_j)^2  \\
  &\phantom{H_{\text{Fen}}}=\frac{L}{2}-\sum_{j=1}^{L}\beta_{j-1}\beta_{j+1} \gamma_{j-1} \gamma_{j}.
  \end{split}
\end{align}

In \eqref{eq:3}, $H_0$ is a constant term added to incorporate supersymmetry and set the energy origin to zero. The non-interacting term \eqref{eq:4} represents the critical Kitaev chain (CKC) \cite{kitaev2001unpaired}, described by a gapless free Majorana CFT with central charge $c=1/2$ at low energies. The interaction term \eqref{eq:5} corresponds to the integrable Fendley model \cite{fendley2019free}, which reduces to a free fermion model under a nonlocal transformation. The Fendley model exhibits extensive ground-state degeneracy (termed superfrustration \cite{huijse2008superfrustration}) and is gapless, although its low-energy behavior cannot be described by a conventional CFT. Importantly, supersymmetry remains unbroken in the ground state, as will be discussed.

Having defined a supersymmetric Majorana fermion model \eqref{eq:Ham} on a one-dimensional periodic chain, let us consider the phase structure.
When $\alpha$ is small, the Hamiltonian $H_{\text{CKC}}$ dominates, driving the system into a gapless phase described by the free Majorana CFT. As shown in \cite{miura2024interacting}, this phase is gapless owing to the Nambu-Goldstone fermion associated with supersymmetry breaking (SUSY SSB).

Ref.~\cite{miura2024interacting} demonstrated that the Hamiltonian becomes frustration-free at $\alpha = 1/2$, and that the exact trivial and topological ground state consists of product states. 
At this point, supersymmetry remains unbroken. As $\alpha$ approaches 1, the system converges to the Fendley model $H_{\text{Fen}}$, which is shown to be gapless with extensive degeneracy. In this paper, we focus on the phase structure in the range $0 \leq \alpha \leq 1/2$, where numerical calculations are feasible.

\subsection{Supersymmetry Breaking Phase Transition}
\subsubsection{Ground state energy and supersymmetry}
\label{Supersymmetry Breaking and Phase Transitions}
Supersymmetry (SUSY) breaking is a key aspect in understanding phase transitions in our model. In this section, we summarize the conditions and implications of SUSY breaking by examining the ground-state energy density, excitation gaps, and critical phenomena.

To define supersymmetry breaking, 
instead of the ground-state energy $E_{\text{g.s.}}$ itself, we use its density
$E_{\text{g.s.}}\qty(\alpha;L)/L$. Following \cite{sannomiya2016supersymmetry,sannomiya2017supersymmetry,sannomiya2019supersymmetry}, we say that SUSY is broken when the ground-state energy density is strictly positive:
\begin{align}
    \label{eq:6}
    \frac{E_{\text{g.s.}}\qty(\alpha;L)}{L}>0.
\end{align}
When $E_{\text{g.s.}}\qty(\alpha;L)/L=0$, SUSY is defined as unbroken. This definition applies to both finite and infinite systems.
We now characterize the phase transition at $\alpha_{\text{c}}$ in terms of physical observables. The SUSY-restoring transition occurs when the ground-state energy density
\begin{equation}
e_{\text{g.s.}}(\alpha) = \lim_{L\to \infty} \frac{E_{\text{g.s.}}\qty(\alpha;L)}{L} \, (\geq 0)
\label{eqn:Egs-density}
\end{equation}
becomes zero.
In the infinite size limit, the definition of SUSY breaking is independent of boundary conditions \cite{miura2024interacting}.

For small $\alpha$, the constant term $H_0$ dominates in \eqref{eq:Ham}, and SUSY is trivially broken. In contrast, at $\alpha=1$, SUSY remains unbroken as shown in Ref.~\cite{fendley2019free}. Using the variational principle and the exact ground-state energy, the critical point $\alpha_{\text{c}}$ for SUSY breaking is bounded as \cite{sannomiya2024spontaneous,miura2024interacting}:
\begin{align}
\label{eq:bound-alpha}
    \qty(0.392\cdots =)\, \frac{\pi}{8}\leq\alpha_{\text{c}}<\hf \; .
\end{align}
From a mean-field approximation with superfields, the critical point is approximated as \cite{miura2024interacting}:
\begin{align}
    \alpha_{\text{c}}^{\text{mf}}=\frac{\pi}{4+\pi}=0.439\cdots \; .
\end{align}
In this paper, we numerically determine the precise value of $\alpha_{\text{c}}$ and investigate the critical behavior of the SUSY breaking transition.

\subsubsection{Excitation gap and supersymmetry}
\label{Excitation Gap in Supersymmetry Breaking Transition}
In the small-$\alpha$ phase ($\alpha < \alpha_{\text{c}}$), broken SUSY imposes specific constraints on the excitation spectrum. In relativistic theories, the supersymmetric counterpart to the Nambu-Goldstone theorem predicts the appearance of a massless fermionic particle, known as the Nambu-Goldstone (NG) fermion, when SUSY is broken \cite{Salam-S-74}. In non-relativistic lattice models like ours, SUSY breaking is often associated with gapless behavior, as observed in previous studies \cite{sannomiya2016supersymmetry,sannomiya2017supersymmetry,sannomiya2019supersymmetry,sannomiyaDron,miura2023supersymmetry,miura2024interacting}. This gapless phase is supported by the presence of gapless Majorana fermion excitations.

On the other hand, in the large-$\alpha$ phase ($\alpha > \alpha_{\text{c}}$), unbroken SUSY does not prevent the opening of a gap. 
The following two possibilities can be considered. One is that the gap opens exactly at $\alpha = \alpha_{\text{c}}$. The other is the persistence of a new gapless phase, where SUSY is restored, even for $\alpha > \alpha_{\text{c}}$. Our numerical results support the former scenario, confirming that the restoration of SUSY and the gap opening occur simultaneously at $\alpha = \alpha_{\text{c}}$.

\section{Numerical Determination of the SUSY Restoration Transition Point $\alpha_{\text{c}}$}
\label{sec:critical-pt-by-numerics}
In this section, we numerically determine the transition point $\alpha_{\text{c}}$ of spontaneous supersymmetry breaking (SUSY SSB) using exact diagonalization and the density matrix renormalization group (DMRG) method \cite{PhysRevLett.69.2863}. We employed the Julia implementation of ITensor \cite{iTensor} for the DMRG computations.
The system under study is fermionic with periodic boundary conditions (PBC) imposed, and the truncation error was set to $10^{-10}$. We performed simulations for system sizes $L=14,16,18,\cdots,50$, starting from random matrix product states.

\subsection{Ground State Energy Density}
\label{subsec:GS-energy-density}

First, we employ Eq.~\eqref{eqn:Egs-density} to numerically identify the critical point
$\alpha_{\text{c}}$, where the ground-state energy density of an infinite system transitions from positive to zero, marking the restoration of supersymmetry. We first calculated the finite-size ground-state energy density $E_{\text{g.s.}}\qty(\alpha;L)/L$ of systems with periodic boundary conditions using DMRG, and then extrapolated it to $L \to \infty$ using the fitting form $E_{\text{g.s.}}\qty(\alpha;L)/L=e_{\text{g.s.}}\qty(\alpha)+a/L^b$ to obtain $e_{\text{g.s.}}\qty(\alpha)$.

\begin{figure}[htb]
  \includegraphics[width=\columnwidth,clip]{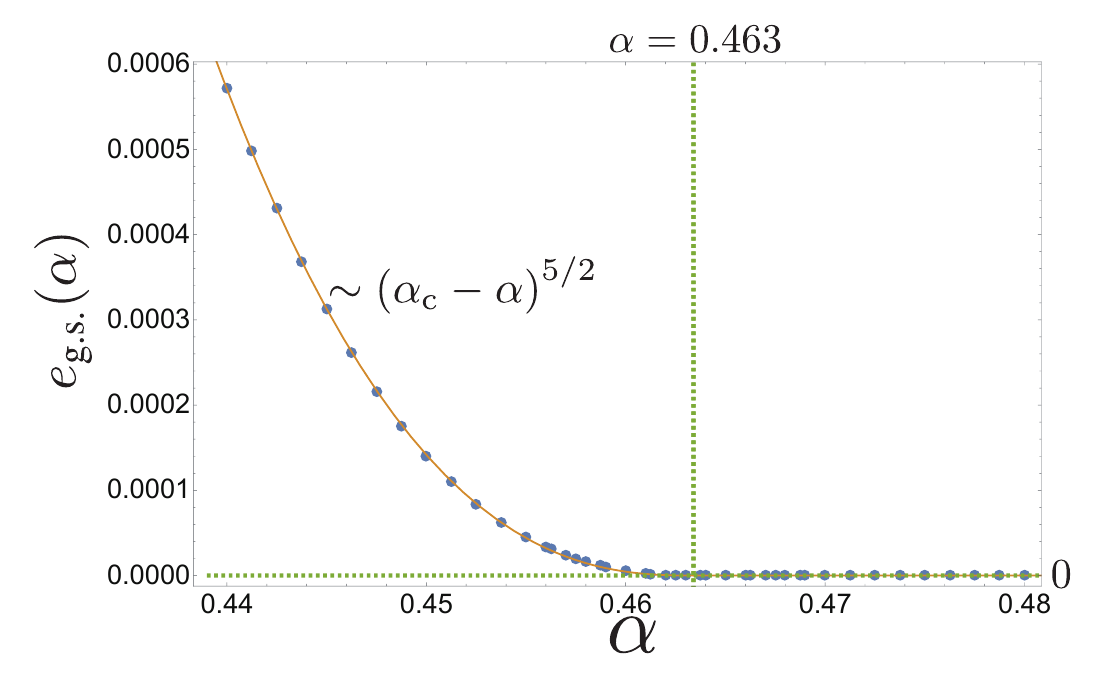}
  \caption{The ground-state energy density $e_{\text{g.s.}}\qty(\alpha)$ as a function of $\alpha$.  
    The dotted-green horizontal line represents $e_{\text{g.s.}}\qty(\alpha)=0$. The data were fitted to the scaling form \eqref{eqn:Edensity-power-law} (orange curve), showing that the energy density vanishes as $e_{\text{g.s.}}\qty(\alpha)\sim \qty(\alpha_{\text{c}}-\alpha)^{5/2}$, indicating the SUSY SSB transition at $\alpha_{\text{c}}=0.463$.  
\label{Fig-energydens}}
\end{figure}

The resulting ground-state energy density $e_{\text{g.s.}}\qty(\alpha)$ as a function of $\alpha$ is plotted in Fig.~\ref{Fig-energydens}. The orange curve represents a power-law fit to the data:
\begin{align}
    e_{\text{g.s.}}\qty(\alpha) = 
\begin{cases} 
a\qty(\alpha_{\text{c}}-\alpha)^{5/2} & \text{for  $\alpha < \alpha_{\text{c}}$}, \\
0 & \text{for  $\alpha_{\text{c}} \leq \alpha$}.
\end{cases}
\label{eqn:Edensity-power-law}
\end{align}
The above-scaling exponent $5/2$ is obtained from the scaling dimension $x_t=6/5$ of the vacancy operator as $2/\qty(2-x_t)=5/2$, with the expectation that the critical point of our model is described by the tricritical Ising (TCI) conformal field theory (CFT) (see Sec.~\ref{sec:behavior_at_transition_point}).
From this fit, we observe that the energy density approaches zero at $\alpha\approx0.463$ and this suggests that the SUSY SSB transition occurs at $\alpha=\alpha_{\text{c}}\approx0.463$, which is consistent with the rigorous bound given in Eq.~\eqref{eq:bound-alpha}.

\subsection{Energy Gap from the Ground State}
\label{subsec:energy_gap_from_gs}
According to the supersymmetric counterpart of the Nambu-Goldstone theorem, the system remains gapless as long as SUSY is spontaneously broken \cite{sannomiya2016supersymmetry,sannomiya2017supersymmetry,sannomiya2019supersymmetry,sannomiyaDron,miura2023supersymmetry,miura2024interacting}. Thus, any gap opening must occur at or above the SUSY SSB transition point $\alpha_{\text{c}}$. In this section, we numerically examine whether SUSY breaking and the gap opening occur simultaneously or at different points $\alpha_{\text{c}}$ and $\alpha'_{\text{c}}$.

\begin{figure}[htbp]
  \includegraphics[width=\columnwidth,clip]{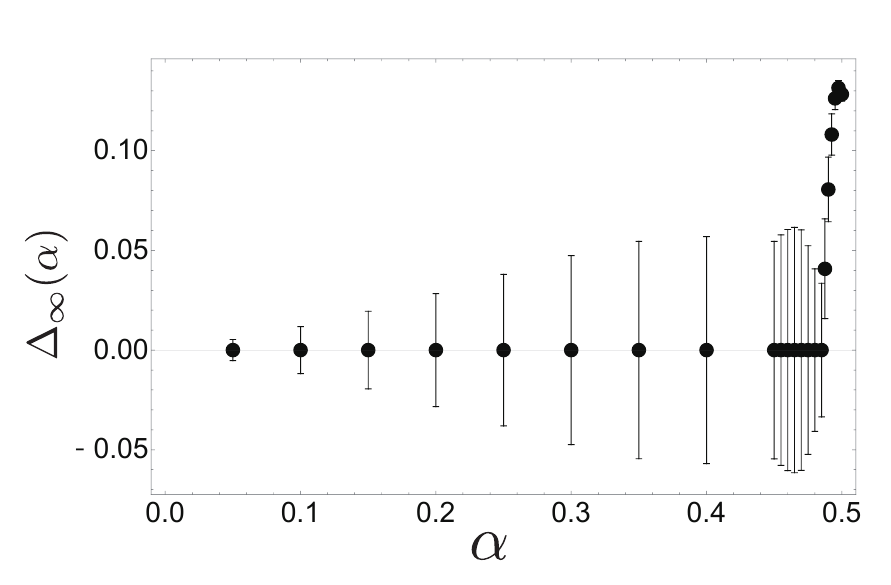}
  \caption{The extrapolated energy gap $\Delta_{\infty}\qty(\alpha)$ as a function of $\alpha$. The gap was obtained by fitting finite-size data from exact diagonalization. The gap becomes significantly finite around $\alpha = 0.485 \sim 0.4875$. Error bars are larger near the critical point $\alpha=\alpha_{\text{c}}\approx0.463$.
  \label{fig:04}}
\end{figure}

To this end, we calculated the infinite-size energy gap $\Delta_{\infty}\qty(\alpha)$ by fitting the finite-size data $\Delta\qty(L;\alpha)$ (obtained from exact diagonalization for periodic systems with $L=4,5,\cdots,15$) to the formula $\Delta\qty(L;\alpha)=\Delta_{\infty} \qty(\alpha)+a/L^b$.
The finite-size gap $\Delta\qty(L;\alpha)$ is defined as the energy difference between the lowest and third-lowest energy levels~\cite{note-gap-definition}.
The extrapolated energy gap $\Delta_{\infty}\qty(\alpha)$ is plotted in Fig.~\ref{fig:04} as a function of $\alpha$.

In supersymmetric quantum systems, SUSY SSB leads to the appearance of the NG fermions, forcing the system to be gapless, as is supported by the single-mode approximation \cite{sannomiya2016supersymmetry,sannomiya2017supersymmetry,sannomiya2019supersymmetry,miura2024interacting}. Furthermore, for $\alpha\ll1$, the Hamiltonian is dominated by $H_{\text{CKC}}$, and a mean-field theory \cite{miura2024interacting} suggest that the low-energy behavior in the region $\alpha<\alpha_{\text{c}}$ is described by the Ising CFT, which is gapless. Our results, as shown in Fig.~\ref{fig:04}, are not contradictory to this scenario though not conclusive, thereby suggesting that the system remains gapless at least for $\alpha \leq \alpha_{\text{c}} \approx 0.463$.

The numerical fit in Fig.~\ref{fig:04} indicates that 
$\Delta_{\infty}\qty(\alpha)$ becomes significantly finite between $\alpha = 0.485 $ and $ 0.4875$. 
Additionally, the error becomes relatively large near $\alpha = \alpha_{\text{c}}\approx0.463$, 
where the ground-state energy density $e_{\text{g.s.}}\qty(\alpha)$ transitions to zero.
Due to the large error, we cannot conclude solely from these data whether the gap-opening point $\alpha'_{\text{c}}$ and the SSB transition point $\alpha_{\text{c}}$ are equal or we have two separate transitions ($\alpha_{\text{c}} < \alpha'_{\text{c}}$) instead.
At least, the value $\alpha'_{\text{c}}=0.485 \sim 0.4875$ does not contradict the general constraint $\alpha_{\text{c}} < \alpha'_{\text{c}}$ required by the NG theorem. 
In the next section, we will argue that $\alpha_{\text{c}}$ and $\alpha'_{\text{c}}$ indeed coincide by carefully analyzing the critical behavior around the SUSY SSB transition.

\section{Critical Behavior at the Phase Transition Point $\alpha_{\text{c}}$}
\label{sec:behavior_at_transition_point}
As discussed above, SUSY is spontaneously broken for sufficiently small $\alpha$, and the system remains gapless due to the presence of the NG fermion. In this region, the term $H_{\text{CKC}}$, which corresponds to the critical Kitaev chain \eqref{eq:4}, dominates the physical behavior, implying that a non-interacting massless Majorana particle serves as the NG fermion. The gapless phase extends at least up to $\alpha=\pi/8$, as suggested by the bound \eqref{eq:bound-alpha}. A mean-field theory predicts \cite{miura2024interacting} that the entire region below the SUSY-restoration transition point $\alpha_{\text{c}}$ is described by the free massless Majorana CFT.

\subsection{Field Theory Description}
\label{sec:perturbed-CFT}
Next, we consider the critical behavior at the point $\alpha = \alpha_{\text{c}}$. Given the underlying microscopic supersymmetry in the model \eqref{eq:4}, the natural candidate for the criticality is the tricritical Ising (TCI) universality class, which is described by the $\mathcal{N}=1$ superconformal field theory with central charge $c = 7/10$ \cite{Friedan-Q-S-SUSY-85,Qiu-86}. This CFT governs the multicritical point between the first-order and second-order phase transitions in the tricritical Ising (or Blume-Capel) model \cite{PhysRevA.4.1071}. The theory contains six primary fields with specific conformal weights: $\mathbf{1}$ ($h_{\mathbf{1}}=0$), $\sigma$ ($h_{\sigma}=3/80$), $\epsilon$ ($h_{\epsilon}=1/10$), $\sigma^{\prime}$ ($h_{\sigma^{\prime}}=7/16$), $t$ ($h_{t}=3/5$), and $G$ ($h_{G}=3/2$). The operator content of the theory, determined by the allowed scaling fields, depends on the boundary conditions, as outlined in Appendix~\ref{low_energy_behavior_in_fermion_TCI}.

The phase structure near the critical point is well captured by the $c=7/10$ TCI CFT perturbed by the relevant $t$-field, corresponding to the vacancy chemical potential \cite{Kastor-M-S-89} (Fig.~\ref{Fig-PhaseDiagram}):
\begin{equation}
\mathcal{S}= \mathcal{S}^{\text{(TCI)}} + g \int\! \rmm{d}^{2}x \, \Phi^{\text{(TCI)}}_{(1,3)}(z,\bar{z})  \; ,
\label{eqn:TCIM-13perturbed}
\end{equation}
where $\mathcal{S}^{\text{(TCI)}}$ is the action of the $c=7/10$ TCI CFT, and $\Phi^{\text{(TCI)}}_{(1,3)}$ is its primary field $t$ with conformal weights $(\bar{h},h)=(3/5,3/5)$. The conservation of parity $(-1)^{F}$ (or the $\mathbb{Z}_{2}$ spin-flip symmetry $\sigma^{x} \to - \sigma^{x}$) allows two relevant perturbations: $\epsilon$ ($h_{\epsilon}=1/10$) and $t$ ($h_{t}=3/5$). However, the exact SUSY of the model \eqref{eq:Ham} prohibits the SUSY-breaking $\epsilon$ from appearing in the effective theory \cite{Kastor-M-S-89}, making Eq.~\eqref{eqn:TCIM-13perturbed} the only symmetry-allowed action for this model.
The coupling constant $g$ describes the deviation from the critical point with $g=0$ at $\alpha=\alpha_{\text{c}}$.

As demonstrated both perturbatively \cite{Zamolodchikov-c-87,Ludwig-C-87} and exactly \cite{Zamolodchikov-TCIM-Ising-91}, there exists an integrable massless renormalization group (RG) flow from the $c=7/10$ TCI fixed point to the $c=1/2$ Ising fixed point when the coupling $g$ is positive. 
Around the infrared Ising ($c=1/2$) fixed point, the RG flow is described by the Ising CFT perturbed by an irrelevant operator \cite{Kastor-M-S-89,Zamolodchikov-TCIM-Ising-91}:
\begin{equation}
\label{Isingperturbation}
\mathcal{S} = \mathcal{S}^{\text{(Ising)}} + g^{\prime} \int\! \rmm{d}^{2}x \, \bar{T}_{\text{Ising}}(\bar{z}) T_{\text{Ising}}(z),
\end{equation}
where $\mathcal{S}^{\text{(Ising)}}$, $\bar{T}_{\text{Ising}}(\bar{z})$, and $T_{\text{Ising}}(z)$ represent the action and energy-momentum tensors of the Ising CFT, respectively. In the weak coupling region $\qty(\alpha \ll 1)$, taking the continuum limit of \eqref{eq:Ham} confirms that the action takes the form of \eqref{Isingperturbation} with $g' \propto \frac{a^{2}\alpha}{1-\alpha}$. 
This implies that the weak-coupling $c=1/2$ Ising phase is stable against the fermion-fermion interaction $H_{\text{F}}$. 
In the fermionic model with periodic boundary conditions, the ground state is two-fold degenerate throughout the Ising phase due to the gapless fermion zero mode. 
In contrast, for negative $g$, the system flows towards a doubly degenerate phase with gapped kink excitations \cite{Zamolodchikov-RSOS-TBA-91,Lepori_2008} and presumably renormalizes eventually to the infinitely massive fixed point with $\alpha=1/2$ (Fig.~\ref{Fig-PhaseDiagram}).

In the model \eqref{eq:Ham}, decreasing (increasing) $\alpha$ from $\alpha_{\text{c}}$ corresponds to positive (negative) $g$ in the perturbed action \eqref{eqn:TCIM-13perturbed}. In the SUSY-unbroken phase $\alpha_{\text{c}} < \alpha \leq \hf$ (near $\alpha = \alpha_{\text{c}}$), the field-theoretical description predicts doubly degenerate ground states with an excitation gap $\Delta_{\infty}(\alpha) \sim (\alpha - \alpha_{\text{c}})^{5/4}$ \cite{Laessig-M-C-91}. Meanwhile, the massless flow toward the $c=1/2$ Ising fixed point for $\alpha < \alpha_{\text{c}}$ is consistent with the observation that the low-energy physics of \eqref{eq:Ham} is dominated by non-interacting massless Majorana fermions. Exact diagonalization shows that the ground-state degeneracy stays at two for $0<\alpha\leq 1/2$ when the fermionic periodic boundary conditions are imposed. The observed degeneracy agrees with the prediction from the $\Phi^{\text{(TCI)}}_{(1,3)}$-perturbed CFT. 

In the following two subsections, we numerically calculate the entanglement entropy and the spectral quantity called the universal gap ratio to compare with the above field-theory predictions for $\alpha \leq \alpha_{\text{c}}$. 

\subsection{Central Charge from Entanglement Entropy}
\label{subsec:behavior_of_EE}
In one-dimensional quantum critical systems, the entanglement entropy $S$ of the ground state depends on the central charge $c$ of the underlying CFT and the size $\ell$ of the subsystem as \cite{HOLZHEY1994443,PhysRevLett.90.227902,PasqualeCalabrese_2004}
\begin{align}
    \label{eqn:Cardi}
    S\qty(\ell;L)\sim\frac{c}{6}\mathcal{A}\ln\qty[\frac{L}{\pi}\sin(\frac{\pi \ell}{L})],
\end{align}
where $\mathcal{A}$ is the number of entanglement cuts ($\mathcal{A}=1$ and $2$ for open and periodic boundary conditions, respectively).

We estimate the central charge from the entanglement entropy following \cite{PhysRevB.84.094410}. Specifically, we calculate the entanglement entropy of the numerically obtained ground state wave function at $\ell = L/2$ and $L/4$ and compute the difference as 
\begin{align}
\label{eq:EEXA}
    \Delta S(L) &\coloneqq S(L/2;L) - S(L/4;L) \notag \\
    &\sim
    \begin{cases}
         \frac{c}{12} \mathcal{A} \ln(2) &\quad \qty(\text{critical phase}), \\
        0 &\quad \qty(\text{gapped phase}). \;
    \end{cases}
\end{align}

\begin{figure}[htb]
  \includegraphics[width=\columnwidth,clip]{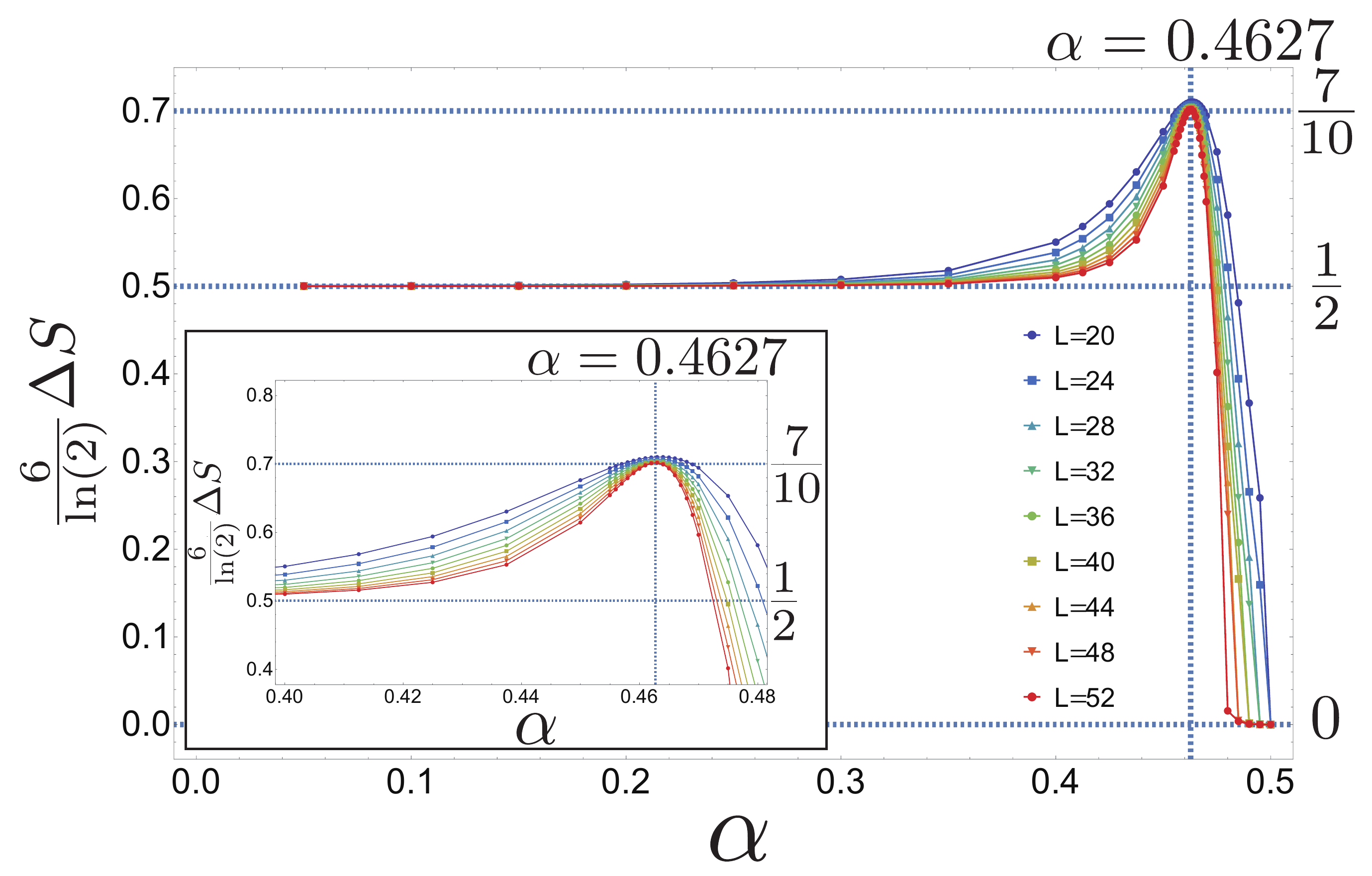}
    \caption{
    The central charge values estimated from the entanglement entropy for each $\alpha$ are shown. In particular, the region around $\alpha \approx \alpha_{\text{c}}$ is magnified and displayed in the lower left corner. As indicated, for $\alpha < 0.4627$, the central charge approaches $1/2$; at $\alpha = 0.4627$, it reaches $7/10$; and for $\alpha > 0.4627$, it approaches zero.
    \label{Fig-a-c-Graph}}
\end{figure}

The numerical calculations were performed using finite DMRG for the spin representation [see (A3)] with periodic boundary conditions, and $\mathcal{A}$ was set to 2. The results of $\Delta S(L)$ [normalized by $\mathcal{A} \ln(2)/12$] for various $L$ are shown in Fig.~\ref{Fig-a-c-Graph}. 
For $\alpha \lessapprox 0.3$, the data for different $L$ collapse onto the $c=1/2$ line indicating the Ising CFT. The plot forms a peak at $\alpha \approx 0.463$, which sharpens as $L$ increases as expected from \eqref{eq:EEXA}. The data collapse now onto the value $c=7/10$ characteristic of the TCI CFT indicating that the transition out of the SUSY broken phase is indeed governed by the TCI CFT.  Additionally, for $0.463 < \alpha \leq 1/2$, the plot approaches zero as $L$ increases. 

By extrapolating the peak position of the interpolated curve for $L$, we estimate the critical point to be $\alpha_{\text{c}} \approx 0.4627$. Combining all these, we conclude that for $\alpha < \alpha_{\text{c}}$, the system is described by the Ising CFT with $c = 1/2$, at $\alpha = \alpha_{\text{c}}$ by the TCI CFT with $c = 7/10$, and for $\alpha > \alpha_{\text{c}}$, by a massive theory.
Good data collapse for $\alpha \lessapprox 0.3$ and $\alpha = \alpha_{\text{c}}$ suggests that the  finite-size effects are small for these values.
\subsection{Universal Gap Ratio}
\label{subsec:univ_gap_ratio}
To further validate the prediction regarding the criticality at $\alpha\leq\alpha_{\text{c}}$, we carry out a spectroscopic analysis. Specifically, we use a quantity called the universal gap ratios \cite{rahmani2015emergent,PhysRevB.92.235123}, which are constructed by combining energies from different sectors or boundary conditions and provide detailed information beyond the central charge.

According to CFT, the energy spectrum of a finite-sized periodic system of length $L$ is closely related to the eigenvalues of the Virasoro generators $L_0$ and $\bar{L}_0$ \cite{Cardy-86,Cardy-86-bc}:
\begin{align}
\label{finiteenergy}
H  = E_0(L) + \frac{2\pi v}{L} \qty(L_0 + \bar{L}_0)  \; ,
\end{align}
where $E_0(L)=e_{\text{g.s.}}\qty(\alpha)L-\frac{\pi v}{6L}c$ is the finite-size ground-state energy, dependent on the central charge $c$, and $v$ is a non-universal velocity parameter. The eigenvalues of $L_0 + \bar{L}_0$ depend on the conformal weights $h$ and $\bar{h}$ of the underlying CFT, and the allowed values of $(h,\bar{h})$ are determined by the boundary conditions (see Appendix~\ref{low_energy_behavior_in_fermion_TCI}). The universal gap ratios \cite{rahmani2015emergent,PhysRevB.92.235123} exploit these relations to serve as fingerprints of the CFT controlling the low-energy behavior.

The model \eqref{eq:Ham} preserves fermion parity $(-1)^{F} = \pm 1$, which leads to four distinct energy spectra depending on the fermion parity (even/odd) and boundary conditions (periodic/antiperiodic). By calculating the energies of the ground and excited states in these four cases, we can form combinations that eliminate non-universal constants like $e_{\text{g.s.}}$ and $v$, yielding universal rational numbers characteristic of the Ising or TCI universality classes.

The lowest two energy eigenvalues for fermion parity even/odd and for periodic (PBC) and antiperiodic (APBC) boundary conditions are summarized in Table~\ref{table:spec-spin-CFT}. These energies are combined into the following universal gap ratios:
\begin{equation}
\begin{split}
& R_{1}=\frac{E_{\text{odd}}^{\text{A},0} (L)- E_{\text{even}}^{\text{A},0}(L)}{E_{\text{even}}^{\text{A},1} (L) - E_{\text{even}}^{\text{A},0} (L)} \, , \quad
R_{2}=\frac{E_{\text{even}}^{\text{P},0} (L) - E_{\text{even}}^{\text{A},0}(L)}{E_{\text{even}}^{\text{A},1}(L) - E_{\text{even}}^{\text{A},0}(L)} \, , \\ 
& R_{3}=\frac{E_{\text{even}}^{\text{P},1}(L) - E_{\text{even}}^{\text{A},0}(L)}{E_{\text{even}}^{\text{A},1} (L)- E_{\text{even}}^{\text{A},0}(L)} \; ,
\end{split}
\label{eqn:def-univ-gap-ratios}
\end{equation}
where A/P denotes antiperiodic/periodic boundary conditions for fermions, 0 (1) represents the ground (first excited) state, and even/odd refers to the fermion parity. For instance, $E_{\text{even}}^{\text{A},0}(L)$ denotes the ground-state energy of the even-parity sector under antiperiodic boundary conditions. If the spectrum assumes the CFT form \eqref{finiteenergy}, the system size $L$ and the non-universal velocity parameter $v$ are eliminated from the ratios, and they take universal values that depend only on the conformal weights of the underlying CFT. The values for the Ising ($c=1/2$) and TCI ($c=7/10$) CFTs are summarized in Table~\ref{table:CFTratio}. 
 
\begin{figure}[H]
    \begin{center}
        \includegraphics[width=\columnwidth,clip]{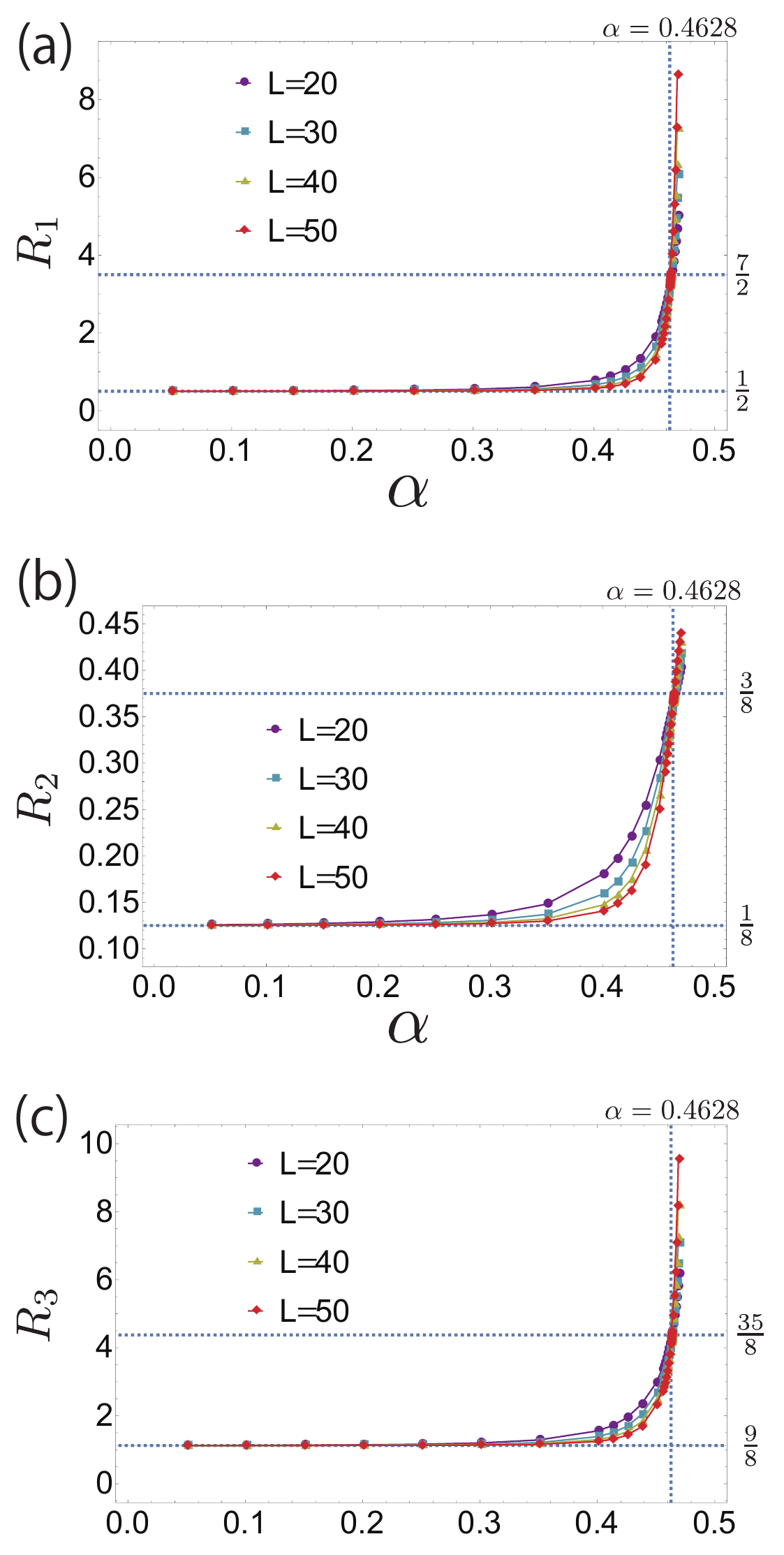}
    \end{center}
    \caption{
        Universal gap ratios (see Table \ref{table:CFTratio}) were calculated from the ground and first excited state energies ($L=20,30,40,50$) using finite DMRG. All approach the TCI values near $\alpha\approx0.463$.
Panels (a), (b), and (c) correspond to the universal ratios
$ R_{1}=\frac{E_{\text{odd}}^{\text{A},0} (L)- E_{\text{even}}^{\text{A},0}(L)}{E_{\text{even}}^{\text{A},1} (L) - E_{\text{even}}^{\text{A},0} (L)},
R_{2}=\frac{E_{\text{even}}^{\text{P},0} (L) - E_{\text{even}}^{\text{A},0}(L)}{E_{\text{even}}^{A,1}(L) - E_{\text{even}}^{\text{A},0}(L)}, 
R_{3}=\frac{E_{\text{even}}^{\text{P},1}(L) - E_{\text{even}}^{\text{A},0}(L)}{E_{\text{even}}^{\text{A},1} (L)- E_{\text{even}}^{\text{A},0}(L)}$, respectively.
The value of $\alpha_{\text{c}}$, obtained by extrapolating $\alpha$ where the gap ratios match the TFI CFT predictions for each $L$, converged to 0.4628 in all cases (a)--(c).
\label{univratio}}
\end{figure}

By comparing these universal gap ratios to the corresponding values obtained numerically, we can identify the CFT governing the low-energy physics \cite{Friedan-Q-S-SUSY-85,rahmani2015emergent}. We calculated $R_{1,2,3}$ for system sizes $L = 16, 18, \cdots, 48, 50$ using finite DMRG. The results are shown in Figs.~\ref{univratio}(a)--\ref{univratio}(c). 
The data for different $L$ collapse onto the Ising and TCI values (see Table~\ref{table:CFTratio}) when $\alpha \lesssim 0.2 \sim 0.3$ and $\alpha \simeq 0.463$, respectively, confirming the conclusions of Sec.~\ref{subsec:behavior_of_EE}. 
To determine the transition point, we first fixed $L$ and then determined $\alpha_{\text{c}}(L)$ at which the interpolated data coincide with the corresponding TCI CFT values. 
By extrapolating $\alpha_{\text{c}}(L)$ to $L \to \infty$, we obtained the transition point $\alpha_{\text{c}} = 0.4628$ in all cases (a), (b), and (c). 
This is consistent with earlier estimates of $\alpha_{\text{c}} \approx 0.463, 0.4627$ from the previous sections~\ref{subsec:GS-energy-density} and \ref{subsec:behavior_of_EE}.

\begin{table}[H]
\centering
\caption{Universal gap ratios \eqref{eqn:def-univ-gap-ratios} for the Ising and tricritical Ising (TCI) CFTs.  
The superscript A/P represents antiperiodic/periodic boundary conditions for fermions, 0/1 represents the ground/first excited state, and 
the subscript even/odd represents even/odd fermion parity (see Appendix~\ref{low_energy_behavior_in_fermion_TCI} 
for details).}
\begin{ruledtabular}  
\begin{tabular}{lcccc}
CFT & $c$ & $R_{1}$ 
& $R_{2}$ 
& $R_{3}$ \\ 
\hline
Ising & $\frac{1}{2}$ & $\frac{1}{2}$ & $\frac{1}{8}$ & $\frac{9}{8}$ \\
TCI & $\frac{7}{10}$ & $\frac{7}{2}$ & $\frac{3}{8}$ & $\frac{35}{8}$ \\ 
\end{tabular}
\end{ruledtabular}  
\label{table:CFTratio}
\end{table}

\section{CONCLUSIONS}
\label{CONCLUSION}
In this paper, we have numerically investigated the ground-state phase diagram of the supersymmetric lattice fermion model given by Eq.~\eqref{eq:Ham} below the solvable frustration-free (F.F.) point $\alpha=1/2$, focusing particularly on spontaneous supersymmetry breaking (SUSY SSB) and the critical behavior near the phase transition out of the SUSY broken phase. This model was originally introduced in Ref.~\onlinecite{miura2024interacting}, where we demonstrated that for small values of the interaction parameter $\alpha$, the system resides in a SUSY-broken phase, while SUSY is restored for the solvable Fendley model $\alpha = 1$.

Through numerical calculations of the ground-state energy density, we first identified the critical point $\alpha_{\text{c}} \approx 0.463$, where SUSY is restored. As the system approaches this point, the energy density vanishes as $e_{\text{g.s.}}\qty(\alpha) \sim \qty(\alpha_c - \alpha)^{5/2}$ which is consistent with the hypothesis that our model is described at low energies by the tricritical Ising (TCI) conformal field theory (CFT) perturbed by the vacancy chemical potential [or the $\Phi_{1,3}^{\text{(TCI)}}$-operator].

Next, we examined the behavior of the energy gap across the SUSY phase transition to see if the gap opening and the SUSY restoration occur simultaneously. According to the supersymmetric analog \cite{miura2024interacting} of the Nambu-Goldstone theorem, the system should remain gapless in the SUSY-broken phase, and any gap opening must occur at or above the critical point $\alpha_c$. Our numerical results indicate that the energy gap starts significantly deviating from zero around $\alpha \approx 0.485\sim0.4875$, which does not contradict the above constraint. Unfortunately, due to the large numerical error near $\alpha_c \approx 0.463$, we cannot definitively conclude whether the gap opening point coincides with the SUSY-restoring point. However, our findings are consistent with the theoretical expectation that the system remains gapless for $\alpha \leq \alpha_c$ and that the gap opens at or above the critical point.

We next estimated the central charge $c$ of the CFT that governs the gapless phase ($\alpha \leq \alpha_{\text{c}}$) by calculating the system-size difference of the von Neumann entanglement entropy, which is designed to give size-independent values ($c$ up to universal constants) when the system is gapless. Our results indicate that for $\alpha < \alpha_c$, the system is described by the Ising CFT with $c = 1/2$. When the critical point $\alpha_{\text{c}}$ is approached, the central charge increases up to the TCI CFT value $c = 7/10$, confirming that the system undergoes a continuous quantum phase transition out of the SUSY broken gapless ($c=1/2$) phase at $\alpha_c$, where the system acquires emergent superconformal symmetry. Moreover, for $\alpha > \alpha_c$, the behavior of the entanglement entropy indicates that the system is gapped and that SUSY SSB and gap opening occur simultaneously at $\alpha=\alpha_{\text{c}}$. 
Again, all these are consistent with the field theoretical predictions based on the TCI CFT perturbed by the relevant $\Phi_{1,3}^{\text{(TCI)}}$-operator. The main results are summarized in Figs.~\ref{Fig-PhaseDiagram} (B) and (C).  

To obtain more insight into the criticality of this transition, we calculated the universal gap ratio in which non-universal quantities like the velocity parameter are eliminated and compared it to the predictions of both the Ising and TCI CFTs. The numerical values at the transition point $\alpha_{\text{c}}$ closely match those of the TCI CFT, providing additional confirmation that the transition belongs to the TCI universality class. 

In this paper, we have reached a fairly precise quantitative understanding of the phase structure of the model \eqref{eq:Ham} below the F.F. point $\alpha =1/2$ with one missing piece being the microscopic understanding of the gapped kink excitations in the gapped phase $\alpha_{\text{c}}<\alpha \leq 1/2$.  
A more important open question remains to figure out how the simple spectrum with doubly degenerate ground states around $\alpha =1/2$ evolves into more complicated ones exhibiting superfrustration at $\alpha =1$. One might speculate that a mechanism similar to that proposed in Ref.~\onlinecite{Chepiga-M-S-21} is at play, whereas further extensive numerics are necessary to answer this question.

\section*{Acknowledgments}
The authors would like to thank T. Ando, P. Fendley, H. Katsura, R. Masui, P. Marra, A. Matsumoto, K. Shimomura, H. Tajima, Y. Nakayama, and A. Ueda for helpful discussions.
U.M. is supported by JST SPRING, Grant No. JPMJSP2110, and JST CREST
Grant No. JPMJCR19T2.
K.T. is supported in part by Japan Society for the Promotion of Science (JSPS) KAKENHI Grant No. 21K03401 
and the IRP project ``Exotic Quantum Matter in Multicomponent Systems (EXQMS)'' from CNRS, France.  
\appendix
\section{Spin Representation of the Hamiltonian via Jordan-Wigner Transformation}
\label{Spin Representation of the Hamiltonian via Jordan-Wigner Transformation}
We present the spin representation of the Hamiltonians \eqref{eq:4} and \eqref{eq:5}, derived via the Jordan-Wigner transformation, which is employed in calculating the entanglement entropy using the density matrix renormalization group (DMRG) method.

For Majorana fermions $\beta_{j}$ and $\gamma_{j}$, the Jordan-Wigner transformation is defined as follows:
\begin{align}
\label{eq:JWtrf}
&\beta_{j} = \qty(-\sz_{1}) \cdots \qty(-\sz_{j-1}) \sx_{j}, \nt
&\gamma_{j} = \qty(-\sz_{1}) \cdots \qty(-\sz_{j-1}) \sy_{j}.
\end{align}
In this representation, the Hamiltonians \eqref{eq:4} and \eqref{eq:5} are transformed into spin variables as follows, excluding the boundary terms:
\begin{equation}
\begin{split}
& H_{\text{CKC}} = \sum_{j} \qty(-\sx_{j}\sx_{j+1} + \sz_{j-1}), \\
& 
H_{\text{F}} = \frac{L}{2} - \sum_{j} \sz_{j-1} \sx_{j} \sx_{j+1}.
\end{split}
\label{eq:spin}
\end{equation}

The omitted boundary terms correspond to a nonlocal fermion parity, which plays a crucial role, especially when considering (anti)periodic boundary conditions in the fermionic system. These boundary effects cannot be neglected in the proper treatment of the system's total parity and should be accounted for in detailed analyses of the low-energy physics.
\section{Low-Energy Spectrum of the Fermion Theory at the Tricritical-Ising and Ising critical points}
\label{low_energy_behavior_in_fermion_TCI}
The model \eqref{eq:Ham} exhibits different (finite-size) spectra depending on the boundary condition. 
As the low-energy part of the finite-size ($L$) spectrum of the lattice model is determined (up to finite-size corrections) 
by the underlying CFTs obeying the same boundary condition, the comparison between the numerical data and the CFT prediction 
provides detailed information on the universality class. 
In this Appendix, we determine the spectrum of the $c=7/10$ TCI CFT for both the fermionic (Majorana) and bosonic (i.e., spin-1/2) realizations. 

The $c=7/10$ TCI CFT \cite{Friedan-Q-S-SUSY-85} has the following six primary fields (with corresponding conformal weights $h$):
\begin{equation}
\label{eq:11}
\begin{split}
& \bmm{1}\, \qty(h=0), \;\; \eps\, \qty(h=1/10) , \;\; t\, \qty(h=3/5), \;\; 
G\, \qty(h=3/2),  \\ 
& \sigma\, \qty(h=3/80), \;\; \sigma^{\prime} \, \qty(h=7/16) \; .
\end{split}
\end{equation}

The most efficient way to find the spectrum under given boundary conditions is to calculate the partition functions (or path integrals) 
for different sets of spatial and temporal boundary conditions:  
\begin{equation}
\begin{split}
Z &= \text{Tr} \, \qty[\be^{- \beta H + i x P}] 
= (q \bar{q})^{-\frac{c}{24}} \, \text{Tr} \, \qty[q^{L_{0}} \bar{q}^{\bar{L}_{0}} ] \\
&= \sum_{h,\bar{h}} N_{h,\bar{h}} \chi_{h} \qty(q) \chi_{\bar{h}} \qty(\bar{q}) \; ,
\end{split}
\label{eqn:partition-fn-general}
\end{equation}
where $q$ and $\bar{q}$ are complex variables that encode the inverse temperature $\beta$ and parameter $x$:
\begin{equation}
q := \be^{2\pi i \left( i \frac{v\beta}{L} + \frac{x}{L} \right)} = \be^{2\pi i \tau} \; , \;  
\bar{q} := \be^{- 2\pi i \left( - i \frac{v\beta}{L} + \frac{x}{L} \right)} = \be^{-2\pi i \bar{\tau}} , 
\end{equation}
and $\chi_{h}(q)$ represents the character of the irreducible representation $(c,h)$ of the Virasoro algebra (for more details, see Ref.~\onlinecite{DiFrancesco-M-S-book}).  
Below, we use shorthand notations like 
\raisebox{-1.7ex}{\includegraphics[scale=0.28]{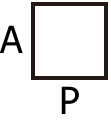}} to specify the spatial and temporal boundary conditions in the path integral representation 
[in this example, periodic (P) and antiperiodic (A) boundary conditions are imposed in the spatial and temporal directions, respectively].  
By expanding Eq.~\eqref{eqn:partition-fn-general} in powers of $q$ with $x$ set to $0$~\cite{note-momentum-resolved}, one can immediately get the finite-size spectrum of the CFT.   

The boundary condition affects the set of non-negative integers $\{ N_{h,\bar{h}} \}$ in Eq.~\eqref{eqn:partition-fn-general}, 
through which the spectrum depends on the boundary condition. 
It is known \cite{Cappelli-I-Z-NP-87,Cappelli-I-Z-CMP-87} that the only partition function  
which is invariant under the {\em full} modular group generated by $S: \, \tau \to - 1/\tau$ and $T: \, \tau \to \tau +1$ is: 
\begin{equation}
\begin{split}
&Z_{\text{S}} \qty(\raisebox{-2.0ex}{\includegraphics[scale=0.3]{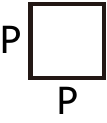}} \; )  \\
&= \abs{\chi_{0}}^{2} + \abs{\chi_{3/80}}^{2} + \abs{\chi_{1/10}}^{2} + \abs{\chi_{7/16}}^{2} + \abs{\chi_{3/5}}^{2} + \abs{\chi_{3/2}}^{2}  \; .
\end{split}
\label{eqn:Zs-PP}
\end{equation}
This is the partition function of the spin (S) model (A3) under periodic boundary conditions in the spatial direction. 
To calculate the partition function under antiperiodic boundary conditions 
$Z_{\text{S}} \qty(\raisebox{-1.8ex}{\includegraphics[scale=0.28]{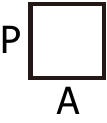}} \; )$, we first compute 
$Z_{\text{S}} \qty(\raisebox{-1.8ex}{\includegraphics[scale=0.28]{./fig09.pdf}} \; )$ by inserting $\FP$ into the trace and 
then apply the modular $S$ transformation to interchange the boundary conditions: 
$Z_{\text{S}} \qty(\raisebox{-1.8ex}{\includegraphics[scale=0.28]{./fig09.pdf}} \; ) \to 
Z_{\text{S}} \qty(\raisebox{-1.8ex}{\includegraphics[scale=0.28]{./fig11.pdf}} \; )$ \cite{Laessig-M-C-91}:
\begin{equation}
\begin{split}
Z_{\text{S}} \qty(\raisebox{-2.0ex}{\includegraphics[scale=0.3]{./fig11.pdf}} \; ) 
=&   \abs{\chi_{3/80}}^{2} + \abs{\chi_{7/16}}^{2} \\ 
& + \chi_{3/5} \bar{\chi}_{1/10}  +  \chi_{1/10} \bar{\chi}_{3/5} + \chi_{3/2} \bar{\chi}_{0} + \chi_{0} \bar{\chi}_{3/2} \; .
\end{split}
\label{eqn:Zs-AP}
\end{equation}

Since the model (6) preserves the fermion parity $(-1)^{F}=\text{even/odd}$, it is convenient to consider the spectrum in the sector 
with a fixed fermion parity.  To this end, we calculate the following {\em parity-resolved} partition functions:
\begin{equation}
\begin{split}
Z^{\text{(even)}}_{\text{S}} \qty(\raisebox{-2.0ex}{\includegraphics[scale=0.3]{./fig10.pdf}} \; )
&=\text{Tr}_{\text{F: APBC}} \qty[\frac{1}{2}\qty(1 + \FP) \be^{-\beta H} ]\\
&=  \abs{\chi_{0}}^{2}  + \abs{\chi_{1/10}}^{2} + \abs{\chi_{3/5}}^{2} + \abs{\chi_{3/2}}^{2}  \\
&= q^{-\frac{7}{120}}
\left\{ 1 + q^{\frac{1}{5}} + 3 q^{\frac{6}{5}} + 2 q^2 + 5  q^{\frac{11}{5}} + \cdots 
\right\} \\
Z^{\text{(odd)}}_{\text{S}} \qty(\raisebox{-2.0ex}{\includegraphics[scale=0.3]{./fig10.pdf}} \; )
&=\text{Tr}_{\text{F: APBC}} \qty[\frac{1}{2}\qty(1 - \FP) \be^{-\beta H} ]\\
&= \abs{\chi_{3/80}}^{2}  + \abs{\chi_{7/16}}^{2}   \\
&= q^{-\frac{7}{120}}
\left\{ q^{\frac{3}{40}} +q^{\frac{7}{8}} +2 q^{\frac{43}{40}} +2 q^{\frac{15}{8}} + \cdots \right\} \; ,
\end{split}
\end{equation}
where we have set $x=0$ for simplicity. 

We can also calculate the parity-resolved version of \eqref{eqn:Zs-AP} in a similar way:
\begin{equation}
\begin{split}
Z^{\text{(even)}}_{\text{S}} \qty(\raisebox{-2.0ex}{\includegraphics[scale=0.3]{./fig11.pdf}} \; )
&= \abs{\chi_{3/80}}^{2}  + \abs{\chi_{7/16}}^{2}  \, \biggl[ = Z^{\text{(odd)}}_{\text{S}} \qty(\raisebox{-2.0ex}{\includegraphics[scale=0.3]{./fig10.pdf}} \; )  \biggr] \\
Z^{\text{(odd)}}_{\text{S}} \qty(\raisebox{-2.0ex}{\includegraphics[scale=0.3]{./fig11.pdf}} \; )
&= \chi_{3/5} \bar{\chi}_{1/10} + \chi_{1/10} \bar{\chi}_{3/5} + \chi_{3/2} \bar{\chi}_{0} + \chi_{0} \bar{\chi}_{3/2}   \\
&= q^{-\frac{7}{120}}
\left\{  2 q^{\frac{7}{10}} +2 q^{\frac{3}{2}} +4 q^{\frac{17}{10}} + 2 q^{\frac{5}{2}}  + \cdots  \right\}\; .
\end{split}
\end{equation}
From there, we can extract the first two lowest energies for each boundary condition and fermion parity, as summarized in Table~\ref{table:spec-spin-CFT}.   
\begin{table}[htb]
\centering
\caption{The lowest two energy levels of the $c=7/10$ TCI CFT in its bosonic (spin) realization 
for given sets of boundary conditions (PBC or APBC) and fermion parity (even or odd).  
The notations $E_{\text{even}}^{\text{A},0}\qty(L)$ and $E_{\text{odd}}^{\text{P},1}\qty(L)$ refer to  
the lowest energy of the parity-even sector for antiperiodic (A) boundary conditions and 
the first excited energy of the parity-odd sector for periodic (P) boundary conditions.}
\begin{ruledtabular}  
\begin{tabular}{lcc}
parity & PBC & APBC  \\ 
\hline
even 
&  
\begin{tabular}{c}
$\tfrac{L}{2\pi v}E_{\text{even}}^{\text{P},0} \qty(L) =0$  \\
$\tfrac{L}{2\pi v}E_{\text{even}}^{\text{P},1}\qty(L) =\tfrac{1}{5}$ 
\end{tabular}
& 
\begin{tabular}{c}
$\tfrac{L}{2\pi v}E_{\text{even}}^{\text{A},0}\qty(L) =\tfrac{3}{40}$  \\
$\tfrac{L}{2\pi v}E_{\text{even}}^{\text{A},1}\qty(L) =\tfrac{7}{8}$ 
\end{tabular} 
\\
\hline
odd & 
\begin{tabular}{c}
$\tfrac{L}{2\pi v}E_{\text{odd}}^{\text{P},0}\qty(L) =\tfrac{3}{40}$  \\
$\tfrac{L}{2\pi v}E_{\text{odd}}^{\text{P},1}\qty(L) =\tfrac{7}{8}$ 
\end{tabular}
& 
\begin{tabular}{c}
$\tfrac{L}{2\pi v}E_{\text{odd}}^{\text{A},0}\qty(L) =\tfrac{7}{10}$  \\
$\tfrac{L}{2\pi v}E_{\text{odd}}^{\text{A},1}\qty(L) =\tfrac{3}{2}$ 
\end{tabular} 
 \\ 
\end{tabular}
\end{ruledtabular}  
\label{table:spec-spin-CFT}
\end{table}

For the Majorana fermion system, we expect that the corresponding CFT contains 
the left and right moving Majorana fermions $\psi$ and $\bar{\psi}$ as well as the supercurrents $G$ and $\bar{G}$:
\begin{equation}
\begin{split}
& \psi \; \qty[\qty(h,\bar{h})=\qty(\tfrac{3}{5},\tfrac{1}{10})] , \; \; \bar{\psi} \; \qty[\qty(h,\bar{h})=\qty(\tfrac{1}{10},\tfrac{3}{5})] ,   \\
& G \; \qty[\qty(h,\bar{h})=\qty(\tfrac{3}{2},0)] , \; \;  \;\;  \bar{G} \; \qty[\qty(h,\bar{h})=\qty(0,\tfrac{3}{2})]  
\end{split}
\end{equation}
as the primary fields.  
We can construct such a theory that is invariant only under a subgroup generated by $S$ and $T^{2}$ \cite{CAPPELLI198782}: 
\begin{equation}
\begin{split}
& Z_{\text{F}} \qty(\raisebox{-2.0ex}{\includegraphics[scale=0.3]{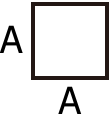}} \; )
=  \abs{\chi_{0}}^{2}  + \abs{\chi_{1/10}}^{2} + \abs{\chi_{3/5}}^{2} + \abs{\chi_{3/2}}^{2} \\
& \phantom{ Z_{\text{F}} \qty(\raisebox{-2.5ex}{\includegraphics[scale=0.35]{./fig12.pdf}} \; ) =  } 
+  \underset{\psi}{ \chi_{3/5} \bar{\chi}_{1/10} }+ \underset{\bar{\psi}}{ \chi_{1/10} \bar{\chi}_{3/5} }
+ \underset{G}{ \chi_{3/2} \bar{\chi}_{0} } + \underset{\bar{G}}{ \chi_{0} \bar{\chi}_{3/2}  }  \\
= & \left( \chi_{0} + \chi_{3/2} \right) \left( \bar{\chi}_{0} + \bar{\chi}_{3/2} \right) 
    + \left( \chi_{1/10} + \chi_{3/5} \right) \left( \bar{\chi}_{1/10} + \bar{\chi}_{3/5} \right)   .
\end{split}
\label{eqn:Zf-AA}
\end{equation}
Fermions are antiperiodic in the imaginary-time direction in the path-integral representation of the trace.  

To calculate the partition function $Z_{\text{F}} \qty(\raisebox{-1.8ex}{\includegraphics[scale=0.28]{./fig09.pdf}} \; )$ 
for periodic (P) boundary conditions in the spatial direction, we first insert $\FP$ into the trace 
of \eqref{eqn:Zf-AA} to change the temporal boundary condition to periodic.   
This flips the signs of the last four terms corresponding to the fermionic fields: 
\begin{equation*}  
\begin{split}
& Z_{\text{F}} \qty(\raisebox{-2.0ex}{\includegraphics[scale=0.3]{./fig12.pdf}} \; ) 
\xrightarrow{\FP}  \\
& \quad 
Z_{\text{F}} \qty(\raisebox{-2.0ex}{\includegraphics[scale=0.3]{./fig11.pdf}} \; )
= \abs{\chi_{0}}^{2}  + \abs{\chi_{1/10}}^{2} + \abs{\chi_{3/5}}^{2} + \abs{\chi_{3/2}}^{2} \\
& \qquad \qquad 
- \chi_{3/5} \bar{\chi}_{1/10} - \chi_{1/10} \bar{\chi}_{3/5} - \chi_{3/2} \bar{\chi}_{0} - \chi_{0} \bar{\chi}_{3/2}  \; .
\end{split}
\end{equation*}
Applying the modular $S$ transformation, we obtain \cite{rahmani2015emergent}:
\begin{equation}
    \label{eq:15}
Z_{\text{F}} \qty(\raisebox{-2.0ex}{\includegraphics[scale=0.3]{./fig09.pdf}} \; )
 =  2 \abs{\chi_{3/80}}^{2}  + 2 \abs{\chi_{7/16}}^{2}  \; .
\end{equation}

We can derive the parity-resolved partition functions in a similar manner:
\begin{subequations}
\begin{align}
\begin{split}
Z^{\text{(even)}}_{\text{F}} \qty(\raisebox{-2.0ex}{\includegraphics[scale=0.3]{./fig12.pdf}} \; )
&= \text{Tr}_{\text{F:A}} \qty[\frac{1}{2}\qty(1 + \FP) \be^{-\beta H}  ]\\
&= \frac{1}{2} Z_{\text{F}} \qty(\raisebox{-2.0ex}{\includegraphics[scale=0.3]{./fig12.pdf}} \; )
+ \frac{1}{2} Z_{\text{F}} \qty(\raisebox{-2.0ex}{\includegraphics[scale=0.3]{./fig11.pdf}} \; )   \\
&= \abs{\chi_{0}}^{2}  + \abs{\chi_{1/10}}^{2} + \abs{\chi_{3/5}}^{2} + \abs{\chi_{3/2}}^{2}  \\
&= Z^{\text{(even)}}_{\text{S}} \qty(\raisebox{-2.0ex}{\includegraphics[scale=0.3]{./fig10.pdf}} \; )  \; ,
\end{split}
\\
\begin{split}
Z^{\text{(odd)}}_{\text{F}} \qty(\raisebox{-2.0ex}{\includegraphics[scale=0.3]{./fig12.pdf}} \; )
&= \text{Tr}_{\text{F:A}} \qty[\frac{1}{2} \qty(1 - \FP) \be^{-\beta H}]   \\
&= \frac{1}{2} Z_{\text{F}} \qty(\raisebox{-2.0ex}{\includegraphics[scale=0.3]{./fig12.pdf}} \; )
- \frac{1}{2} Z_{\text{F}} \qty(\raisebox{-2.0ex}{\includegraphics[scale=0.3]{./fig11.pdf}} \; )   \\
&= \chi_{3/5} \bar{\chi}_{1/10} + \chi_{1/10} \bar{\chi}_{3/5} + \chi_{3/2} \bar{\chi}_{0} + \chi_{0} \bar{\chi}_{3/2} \\
&=  Z^{\text{(odd)}}_{\text{S}} \qty(\raisebox{-2.0ex}{\includegraphics[scale=0.3]{./fig11.pdf}} \; )  \; ,
\end{split}
\end{align}
where $\text{Tr}_{\text{F:A}}$ represents the trace over fermions with antiperiodic boundary conditions.  
Similarly, for periodic boundary conditions, we obtain:
\begin{equation}
\begin{split}
& Z^{\text{(even)}}_{\text{F}} \qty(\raisebox{-2.0ex}{\includegraphics[scale=0.3]{./fig09.pdf}} \; )
= Z^{\text{(odd)}}_{\text{F}} \qty(\raisebox{-2.0ex}{\includegraphics[scale=0.3]{./fig09.pdf}} \; ) \\
&= \frac{1}{2} Z_{\text{F}} \qty(\raisebox{-2.0ex}{\includegraphics[scale=0.3]{./fig09.pdf}} \; )
= \abs{\chi_{3/80}}^{2}  + \abs{\chi_{7/16}}^{2}  \\
&= Z^{\text{(odd)}}_{\text{S}} \qty(\raisebox{-2.0ex}{\includegraphics[scale=0.3]{./fig10.pdf}} \; ) 
= Z^{\text{(even)}}_{\text{S}} \qty(\raisebox{-2.0ex}{\includegraphics[scale=0.3]{./fig11.pdf}} \; )   \; .
\end{split}
\end{equation}
\end{subequations}
In the periodic case, the partition functions are identical for parity even and odd due to the presence of zero modes $G_{0}$ and $\bar{G}_{0}$.   

Using the above partition functions, we can read off the low-energy spectrum of the interacting Majorana fermion at the $c=7/10$ 
critical point, as shown in Table~\ref{table:spec-fermion-CFT}.   
\begin{table}[htb]
\centering
\caption{The lowest two energy levels of the $c=7/10$ TCI CFT in its fermionic realization 
for given sets of boundary conditions (PBC or APBC) and fermion parity (even or odd).  
The notations $E_{\text{even}}^{\text{A},0}\qty(L)$ and $E_{\text{odd}}^{\text{P},1}\qty(L)$ refer to  
the lowest energy of the parity-even sector for antiperiodic (A) boundary conditions and 
the first excited energy of the parity-odd sector for periodic (P) boundary conditions.}
\begin{ruledtabular}  
\begin{tabular}{lcc}
parity & PBC & APBC  \\ 
\hline
even 
&  
\begin{tabular}{c}
$\tfrac{L}{2\pi v}E_{\text{even}}^{\text{P},0} \qty(L) =\tfrac{3}{40}$  \\
$\tfrac{L}{2\pi v}E_{\text{even}}^{\text{P},1} \qty(L) =\tfrac{7}{8}$ 
\end{tabular}
& 
\begin{tabular}{c}
$\tfrac{L}{2\pi v}E_{\text{even}}^{\text{A},0}\qty(L) =0$  \\
$\tfrac{L}{2\pi v}E_{\text{even}}^{\text{A},1}\qty(L)=\tfrac{1}{5}$ 
\end{tabular} 
\\
\hline
odd & 
\begin{tabular}{c}
$\tfrac{L}{2\pi v}E_{\text{odd}}^{\text{P},0}\qty(L) =\tfrac{3}{40}$  \\
$\tfrac{L}{2\pi v}E_{\text{odd}}^{\text{P},1}\qty(L) =\tfrac{7}{8}$ 
\end{tabular}
& 
\begin{tabular}{c}
$\tfrac{L}{2\pi v}E_{\text{odd}}^{\text{A},0}\qty(L) =\tfrac{7}{10}$  \\
$\tfrac{L}{2\pi v}E_{\text{odd}}^{\text{A},1}\qty(L) =\tfrac{3}{2}$ 
\end{tabular} 
 \\ 
\end{tabular}
\end{ruledtabular}  
\label{table:spec-fermion-CFT}
\end{table}

\clearpage

Finally, we provide similar results for the $c=1/2$ Ising CFT in Table~\ref{table:spec-fermion-Ising-CFT}. These results can be extracted from the following parity-resolved partition functions:
\begin{equation}
\begin{split}
& Z^{\text{(even)}}_{\text{F}} \qty(\raisebox{-2.0ex}{\includegraphics[scale=0.3]{./fig12.pdf}} \; )
= \abs{\chi_{0}}^{2}  + \abs{\chi_{1/2}}^{2} 
= q^{-\frac{1}{24}} \left\{ 1 + q + 4  q^2+\cdots \right\} \\
& Z^{\text{(odd)}}_{\text{F}} \qty(\raisebox{-2.0ex}{\includegraphics[scale=0.3]{./fig12.pdf}} \; )
= \chi_{1/2}\bar{\chi}_{0}  + \chi_{0}\bar{\chi}_{1/2} \\
& \phantom{ Z^{\text{(odd)}}_{\text{F}} \qty(\raisebox{-2.0ex}{\includegraphics[scale=0.3]{./fig12.pdf}} \; ) }
= q^{-\frac{1}{24}} \left\{ 2q^{1/2} + 2 q^{3/2} + \cdots \right\} \\
& Z^{\text{(even)}}_{\text{F}} \qty(\raisebox{-2.0ex}{\includegraphics[scale=0.3]{./fig09.pdf}} \; )
= Z^{\text{(odd)}}_{\text{F}} \qty(\raisebox{-2.0ex}{\includegraphics[scale=0.3]{./fig09.pdf}} \; )  \\
& \phantom{Z^{\text{(even)}}_{\text{F}} \qty(\raisebox{-2.0ex}{\includegraphics[scale=0.3]{./fig09.pdf}} \; )} = \abs{\chi_{1/16}}^{2} 
= q^{-\frac{1}{24}} \left\{ q^{1/8} + 2  q^{9/8} + \cdots \right\} \; .
\end{split}
\end{equation}
These universal ratios, along with those 
derived for the TCI CFT, help us determine the universality classes of the critical phases.  

\begin{table}[htb]
\centering
\caption{The lowest two energy levels of the $c=1/2$ Ising CFT in its fermionic realization 
for given sets of boundary conditions (PBC or APBC) and fermion parity (even or odd).  
The notations $E_{\text{even}}^{\text{A},0}\qty(L)$ and $E_{\text{odd}}^{\text{P},1}\qty(L)$ refer to  
the lowest energy of the parity-even sector for antiperiodic (A) boundary conditions and 
the first excited energy of the parity-odd sector for periodic (P) boundary conditions.}
\begin{ruledtabular}  
\begin{tabular}{lcc}
parity & PBC & APBC  \\ 
\hline
even 
&  
\begin{tabular}{c}
$\tfrac{L}{2\pi v}E_{\text{even}}^{\text{P},0} \qty(L) =\tfrac{1}{8}$  \\
$\tfrac{L}{2\pi v}E_{\text{even}}^{\text{P},1} \qty(L) =\tfrac{9}{8}$ 
\end{tabular}
& 
\begin{tabular}{c}
$\tfrac{L}{2\pi v}E_{\text{even}}^{\text{A},0}\qty(L) =0$  \\
$\tfrac{L}{2\pi v}E_{\text{even}}^{\text{A},1}\qty(L) =1$ 
\end{tabular} 
\\
\hline
odd & 
\begin{tabular}{c}
$\tfrac{L}{2\pi v}E_{\text{odd}}^{\text{P},0}\qty(L) =\tfrac{1}{8}$  \\
$\tfrac{L}{2\pi v}E_{\text{odd}}^{\text{P},1}\qty(L) =\tfrac{9}{8}$ 
\end{tabular}
& 
\begin{tabular}{c}
$\tfrac{L}{2\pi v}E_{\text{odd}}^{\text{A},0}\qty(L) =\tfrac{1}{2}$  \\
$\tfrac{L}{2\pi v}E_{\text{odd}}^{\text{A},1}\qty(L) =\tfrac{3}{2}$ 
\end{tabular} 
 \\ 
\end{tabular}
\end{ruledtabular}  
\label{table:spec-fermion-Ising-CFT}
\end{table}

\clearpage

%


\begin{thebibliography}{69}%
\makeatletter
\providecommand \@ifxundefined [1]{%
 \@ifx{#1\undefined}
}%
\providecommand \@ifnum [1]{%
 \ifnum #1\expandafter \@firstoftwo
 \else \expandafter \@secondoftwo
 \fi
}%
\providecommand \@ifx [1]{%
 \ifx #1\expandafter \@firstoftwo
 \else \expandafter \@secondoftwo
 \fi
}%
\providecommand \natexlab [1]{#1}%
\providecommand \enquote  [1]{``#1''}%
\providecommand \bibnamefont  [1]{#1}%
\providecommand \bibfnamefont [1]{#1}%
\providecommand \citenamefont [1]{#1}%
\providecommand \href@noop [0]{\@secondoftwo}%
\providecommand \href [0]{\begingroup \@sanitize@url \@href}%
\providecommand \@href[1]{\@@startlink{#1}\@@href}%
\providecommand \@@href[1]{\endgroup#1\@@endlink}%
\providecommand \@sanitize@url [0]{\catcode `\\12\catcode `\$12\catcode `\&12\catcode `\#12\catcode `\^12\catcode `\_12\catcode `\%12\relax}%
\providecommand \@@startlink[1]{}%
\providecommand \@@endlink[0]{}%
\providecommand \url  [0]{\begingroup\@sanitize@url \@url }%
\providecommand \@url [1]{\endgroup\@href {#1}{\urlprefix }}%
\providecommand \urlprefix  [0]{URL }%
\providecommand \Eprint [0]{\href }%
\providecommand \doibase [0]{https://doi.org/}%
\providecommand \selectlanguage [0]{\@gobble}%
\providecommand \bibinfo  [0]{\@secondoftwo}%
\providecommand \bibfield  [0]{\@secondoftwo}%
\providecommand \translation [1]{[#1]}%
\providecommand \BibitemOpen [0]{}%
\providecommand \bibitemStop [0]{}%
\providecommand \bibitemNoStop [0]{.\EOS\space}%
\providecommand \EOS [0]{\spacefactor3000\relax}%
\providecommand \BibitemShut  [1]{\csname bibitem#1\endcsname}%
\let\auto@bib@innerbib\@empty
\bibitem [{\citenamefont {Wess}\ and\ \citenamefont {Zumino}(1974)}]{WESS197439}%
  \BibitemOpen
  \bibfield  {author} {\bibinfo {author} {\bibfnamefont {J.}~\bibnamefont {Wess}}\ and\ \bibinfo {author} {\bibfnamefont {B.}~\bibnamefont {Zumino}},\ }\bibfield  {title} {\bibinfo {title} {Supergauge transformations in four dimensions},\ }\href {https://doi.org/https://doi.org/10.1016/0550-3213(74)90355-1} {\bibfield  {journal} {\bibinfo  {journal} {Nucl. Phys. B}\ }\textbf {\bibinfo {volume} {70}},\ \bibinfo {pages} {39} (\bibinfo {year} {1974})}\BibitemShut {NoStop}%
\bibitem [{\citenamefont {Witten}(1981)}]{witten1981dynamical}%
  \BibitemOpen
  \bibfield  {author} {\bibinfo {author} {\bibfnamefont {E.}~\bibnamefont {Witten}},\ }\bibfield  {title} {\bibinfo {title} {Dynamical breaking of supersymmetry},\ }\href {https://www.sciencedirect.com/science/article/abs/pii/0550321381900067} {\bibfield  {journal} {\bibinfo  {journal} {Nucl. Phys. B}\ }\textbf {\bibinfo {volume} {188}},\ \bibinfo {pages} {513} (\bibinfo {year} {1981})}\BibitemShut {NoStop}%
\bibitem [{\citenamefont {Witten}(1982)}]{witten1982constraints}%
  \BibitemOpen
  \bibfield  {author} {\bibinfo {author} {\bibfnamefont {E.}~\bibnamefont {Witten}},\ }\bibfield  {title} {\bibinfo {title} {Constraints on supersymmetry breaking},\ }\href {https://www.sciencedirect.com/science/article/abs/pii/0550321382900712} {\bibfield  {journal} {\bibinfo  {journal} {Nucl. Phys. B}\ }\textbf {\bibinfo {volume} {202}},\ \bibinfo {pages} {253} (\bibinfo {year} {1982})}\BibitemShut {NoStop}%
\bibitem [{\citenamefont {Weinberg}(1976)}]{weinberg1976implications}%
  \BibitemOpen
  \bibfield  {author} {\bibinfo {author} {\bibfnamefont {S.}~\bibnamefont {Weinberg}},\ }\bibfield  {title} {\bibinfo {title} {Implications of dynamical symmetry breaking},\ }\href {https://journals.aps.org/prd/abstract/10.1103/PhysRevD.13.974} {\bibfield  {journal} {\bibinfo  {journal} {Phys. Rev. D}\ }\textbf {\bibinfo {volume} {13}},\ \bibinfo {pages} {974} (\bibinfo {year} {1976})}\BibitemShut {NoStop}%
\bibitem [{\citenamefont {Gildener}(1976)}]{PhysRevD.14.1667}%
  \BibitemOpen
  \bibfield  {author} {\bibinfo {author} {\bibfnamefont {E.}~\bibnamefont {Gildener}},\ }\bibfield  {title} {\bibinfo {title} {Gauge-symmetry hierarchies},\ }\href {https://doi.org/10.1103/PhysRevD.14.1667} {\bibfield  {journal} {\bibinfo  {journal} {Phys. Rev. D}\ }\textbf {\bibinfo {volume} {14}},\ \bibinfo {pages} {1667} (\bibinfo {year} {1976})}\BibitemShut {NoStop}%
\bibitem [{\citenamefont {Parisi}\ and\ \citenamefont {Sourlas}(1979)}]{parisi1979random}%
  \BibitemOpen
  \bibfield  {author} {\bibinfo {author} {\bibfnamefont {G.}~\bibnamefont {Parisi}}\ and\ \bibinfo {author} {\bibfnamefont {N.}~\bibnamefont {Sourlas}},\ }\bibfield  {title} {\bibinfo {title} {Random magnetic fields, supersymmetry, and negative dimensions},\ }\href {https://journals.aps.org/prl/abstract/10.1103/PhysRevLett.43.744} {\bibfield  {journal} {\bibinfo  {journal} {Phys. Rev. Lett.}\ }\textbf {\bibinfo {volume} {43}},\ \bibinfo {pages} {744} (\bibinfo {year} {1979})}\BibitemShut {NoStop}%
\bibitem [{\citenamefont {Efetov}(1996)}]{Efetov-book-96}%
  \BibitemOpen
  \bibfield  {author} {\bibinfo {author} {\bibfnamefont {K.}~\bibnamefont {Efetov}},\ }\href@noop {} {\emph {\bibinfo {title} {Supersymmetry in Disorder and Chaos}}}\ (\bibinfo  {publisher} {Cambridge University Press},\ \bibinfo {year} {1996})\BibitemShut {NoStop}%
\bibitem [{\citenamefont {Wiegmann}(1988)}]{PhysRevLett.60.821}%
  \BibitemOpen
  \bibfield  {author} {\bibinfo {author} {\bibfnamefont {P.~B.}\ \bibnamefont {Wiegmann}},\ }\bibfield  {title} {\bibinfo {title} {Superconductivity in strongly correlated electronic systems and confinement versus deconfinement phenomenon},\ }\href {https://doi.org/10.1103/PhysRevLett.60.821} {\bibfield  {journal} {\bibinfo  {journal} {Phys. Rev. Lett.}\ }\textbf {\bibinfo {volume} {60}},\ \bibinfo {pages} {821} (\bibinfo {year} {1988})}\BibitemShut {NoStop}%
\bibitem [{\citenamefont {Bares}\ and\ \citenamefont {Blatter}(1990)}]{PhysRevLett.64.2567}%
  \BibitemOpen
  \bibfield  {author} {\bibinfo {author} {\bibfnamefont {P.~A.}\ \bibnamefont {Bares}}\ and\ \bibinfo {author} {\bibfnamefont {G.}~\bibnamefont {Blatter}},\ }\bibfield  {title} {\bibinfo {title} {Supersymmetric $t$-{$J$} model in one dimension: Separation of spin and charge},\ }\href {https://doi.org/10.1103/PhysRevLett.64.2567} {\bibfield  {journal} {\bibinfo  {journal} {Phys. Rev. Lett.}\ }\textbf {\bibinfo {volume} {64}},\ \bibinfo {pages} {2567} (\bibinfo {year} {1990})}\BibitemShut {NoStop}%
\bibitem [{\citenamefont {Friedan}\ \emph {et~al.}(1985)\citenamefont {Friedan}, \citenamefont {Qiu},\ and\ \citenamefont {Shenker}}]{Friedan-Q-S-SUSY-85}%
  \BibitemOpen
  \bibfield  {author} {\bibinfo {author} {\bibfnamefont {D.}~\bibnamefont {Friedan}}, \bibinfo {author} {\bibfnamefont {Z.}~\bibnamefont {Qiu}},\ and\ \bibinfo {author} {\bibfnamefont {S.}~\bibnamefont {Shenker}},\ }\bibfield  {title} {\bibinfo {title} {Superconformal invariance in two dimensions and the tricritical {I}sing model},\ }\href {https://doi.org/10.1016/0370-2693(85)90819-6} {\bibfield  {journal} {\bibinfo  {journal} {Phys. Lett. B}\ }\textbf {\bibinfo {volume} {151}},\ \bibinfo {pages} {37} (\bibinfo {year} {1985})}\BibitemShut {NoStop}%
\bibitem [{\citenamefont {Qiu}(1986)}]{Qiu-86}%
  \BibitemOpen
  \bibfield  {author} {\bibinfo {author} {\bibfnamefont {Z.}~\bibnamefont {Qiu}},\ }\bibfield  {title} {\bibinfo {title} {Supersymmetry, two-dimensional critical phenomena and the tricritical {I}sing model},\ }\href {https://doi.org/10.1016/0550-3213(86)90553-5} {\bibfield  {journal} {\bibinfo  {journal} {Nucl. Phys. B}\ }\textbf {\bibinfo {volume} {270}},\ \bibinfo {pages} {205} (\bibinfo {year} {1986})}\BibitemShut {NoStop}%
\bibitem [{\citenamefont {Yu}\ and\ \citenamefont {Yang}(2008)}]{yu2008supersymmetry}%
  \BibitemOpen
  \bibfield  {author} {\bibinfo {author} {\bibfnamefont {Y.}~\bibnamefont {Yu}}\ and\ \bibinfo {author} {\bibfnamefont {K.}~\bibnamefont {Yang}},\ }\bibfield  {title} {\bibinfo {title} {Supersymmetry and the {G}oldstino-like mode in {B}ose-{F}ermi mixtures},\ }\href {https://journals.aps.org/prl/abstract/10.1103/PhysRevLett.100.090404} {\bibfield  {journal} {\bibinfo  {journal} {Phys. Rev. Lett.}\ }\textbf {\bibinfo {volume} {100}},\ \bibinfo {pages} {090404} (\bibinfo {year} {2008})}\BibitemShut {NoStop}%
\bibitem [{\citenamefont {Yu}\ and\ \citenamefont {Yang}(2010)}]{yu2010simulating}%
  \BibitemOpen
  \bibfield  {author} {\bibinfo {author} {\bibfnamefont {Y.}~\bibnamefont {Yu}}\ and\ \bibinfo {author} {\bibfnamefont {K.}~\bibnamefont {Yang}},\ }\bibfield  {title} {\bibinfo {title} {Simulating the {W}ess-{Z}umino supersymmetry model in optical lattices},\ }\href {https://journals.aps.org/prl/abstract/10.1103/PhysRevLett.105.150605} {\bibfield  {journal} {\bibinfo  {journal} {Phys. Rev. Lett.}\ }\textbf {\bibinfo {volume} {105}},\ \bibinfo {pages} {150605} (\bibinfo {year} {2010})}\BibitemShut {NoStop}%
\bibitem [{\citenamefont {Ulrich}\ \emph {et~al.}(2014)\citenamefont {Ulrich}, \citenamefont {Adagideli}, \citenamefont {Schuricht},\ and\ \citenamefont {Hassler}}]{PhysRevB.90.075408}%
  \BibitemOpen
  \bibfield  {author} {\bibinfo {author} {\bibfnamefont {J.}~\bibnamefont {Ulrich}}, \bibinfo {author} {\bibfnamefont {I.}~\bibnamefont {Adagideli}}, \bibinfo {author} {\bibfnamefont {D.}~\bibnamefont {Schuricht}},\ and\ \bibinfo {author} {\bibfnamefont {F.}~\bibnamefont {Hassler}},\ }\bibfield  {title} {\bibinfo {title} {Supersymmetry in the majorana {C}ooper-pair box},\ }\href {https://doi.org/10.1103/PhysRevB.90.075408} {\bibfield  {journal} {\bibinfo  {journal} {Phys. Rev. B}\ }\textbf {\bibinfo {volume} {90}},\ \bibinfo {pages} {075408} (\bibinfo {year} {2014})}\BibitemShut {NoStop}%
\bibitem [{\citenamefont {Ebisu}\ \emph {et~al.}(2019)\citenamefont {Ebisu}, \citenamefont {Sagi},\ and\ \citenamefont {Oreg}}]{PhysRevLett.123.026401}%
  \BibitemOpen
  \bibfield  {author} {\bibinfo {author} {\bibfnamefont {H.}~\bibnamefont {Ebisu}}, \bibinfo {author} {\bibfnamefont {E.}~\bibnamefont {Sagi}},\ and\ \bibinfo {author} {\bibfnamefont {Y.}~\bibnamefont {Oreg}},\ }\bibfield  {title} {\bibinfo {title} {Supersymmetry in the insulating phase of a chain of majorana cooper pair boxes},\ }\href {https://doi.org/10.1103/PhysRevLett.123.026401} {\bibfield  {journal} {\bibinfo  {journal} {Phys. Rev. Lett.}\ }\textbf {\bibinfo {volume} {123}},\ \bibinfo {pages} {026401} (\bibinfo {year} {2019})}\BibitemShut {NoStop}%
\bibitem [{\citenamefont {Zhang}\ \emph {et~al.}(2024)\citenamefont {Zhang}, \citenamefont {Guo}, \citenamefont {Tajima},\ and\ \citenamefont {Liang}}]{PhysRevB.110.064512}%
  \BibitemOpen
  \bibfield  {author} {\bibinfo {author} {\bibfnamefont {T.}~\bibnamefont {Zhang}}, \bibinfo {author} {\bibfnamefont {Y.}~\bibnamefont {Guo}}, \bibinfo {author} {\bibfnamefont {H.}~\bibnamefont {Tajima}},\ and\ \bibinfo {author} {\bibfnamefont {H.}~\bibnamefont {Liang}},\ }\bibfield  {title} {\bibinfo {title} {Probing the goldstino excitation through tunneling transport in a {B}ose-{F}ermi mixture with explicitly broken supersymmetry},\ }\href {https://doi.org/10.1103/PhysRevB.110.064512} {\bibfield  {journal} {\bibinfo  {journal} {Phys. Rev. B}\ }\textbf {\bibinfo {volume} {110}},\ \bibinfo {pages} {064512} (\bibinfo {year} {2024})}\BibitemShut {NoStop}%
\bibitem [{\citenamefont {Nicolai}(1976)}]{nicolai1976supersymmetry}%
  \BibitemOpen
  \bibfield  {author} {\bibinfo {author} {\bibfnamefont {H.}~\bibnamefont {Nicolai}},\ }\bibfield  {title} {\bibinfo {title} {Supersymmetry and spin systems},\ }\href {https://iopscience.iop.org/article/10.1088/0305-4470/9/9/010} {\bibfield  {journal} {\bibinfo  {journal} {J. Phys. A: Math. Gen.}\ }\textbf {\bibinfo {volume} {9}},\ \bibinfo {pages} {1497} (\bibinfo {year} {1976})}\BibitemShut {NoStop}%
\bibitem [{\citenamefont {Nicolai}(1977)}]{nicolai1977extensions}%
  \BibitemOpen
  \bibfield  {author} {\bibinfo {author} {\bibfnamefont {H.}~\bibnamefont {Nicolai}},\ }\bibfield  {title} {\bibinfo {title} {Extensions of supersymmetric spin systems},\ }\href {https://iopscience.iop.org/article/10.1088/0305-4470/10/12/022} {\bibfield  {journal} {\bibinfo  {journal} {J. Phys. A: Math. Gen.}\ }\textbf {\bibinfo {volume} {10}},\ \bibinfo {pages} {2143} (\bibinfo {year} {1977})}\BibitemShut {NoStop}%
\bibitem [{\citenamefont {Fendley}\ \emph {et~al.}(2003{\natexlab{a}})\citenamefont {Fendley}, \citenamefont {Schoutens},\ and\ \citenamefont {de~Boer}}]{PhysRevLett.90.120402}%
  \BibitemOpen
  \bibfield  {author} {\bibinfo {author} {\bibfnamefont {P.}~\bibnamefont {Fendley}}, \bibinfo {author} {\bibfnamefont {K.}~\bibnamefont {Schoutens}},\ and\ \bibinfo {author} {\bibfnamefont {J.}~\bibnamefont {de~Boer}},\ }\bibfield  {title} {\bibinfo {title} {Lattice models with $\mathcal{N}=2$ supersymmetry},\ }\href {https://doi.org/10.1103/PhysRevLett.90.120402} {\bibfield  {journal} {\bibinfo  {journal} {Phys. Rev. Lett.}\ }\textbf {\bibinfo {volume} {90}},\ \bibinfo {pages} {120402} (\bibinfo {year} {2003}{\natexlab{a}})}\BibitemShut {NoStop}%
\bibitem [{\citenamefont {Fendley}\ \emph {et~al.}(2003{\natexlab{b}})\citenamefont {Fendley}, \citenamefont {Nienhuis},\ and\ \citenamefont {Schoutens}}]{fendley2003lattice}%
  \BibitemOpen
  \bibfield  {author} {\bibinfo {author} {\bibfnamefont {P.}~\bibnamefont {Fendley}}, \bibinfo {author} {\bibfnamefont {B.}~\bibnamefont {Nienhuis}},\ and\ \bibinfo {author} {\bibfnamefont {K.}~\bibnamefont {Schoutens}},\ }\bibfield  {title} {\bibinfo {title} {Lattice fermion models with supersymmetry},\ }\href {https://iopscience.iop.org/article/10.1088/0305-4470/36/50/004} {\bibfield  {journal} {\bibinfo  {journal} {J. Phys. A: Math. Gen.}\ }\textbf {\bibinfo {volume} {36}},\ \bibinfo {pages} {12399} (\bibinfo {year} {2003}{\natexlab{b}})}\BibitemShut {NoStop}%
\bibitem [{\citenamefont {Fendley}\ and\ \citenamefont {Schoutens}(2005)}]{fendley2005exact}%
  \BibitemOpen
  \bibfield  {author} {\bibinfo {author} {\bibfnamefont {P.}~\bibnamefont {Fendley}}\ and\ \bibinfo {author} {\bibfnamefont {K.}~\bibnamefont {Schoutens}},\ }\bibfield  {title} {\bibinfo {title} {Exact results for strongly correlated fermions in 2+1 dimensions},\ }\href {https://journals.aps.org/prl/abstract/10.1103/PhysRevLett.95.046403} {\bibfield  {journal} {\bibinfo  {journal} {Phys. Rev. Lett.}\ }\textbf {\bibinfo {volume} {95}},\ \bibinfo {pages} {046403} (\bibinfo {year} {2005})}\BibitemShut {NoStop}%
\bibitem [{\citenamefont {Huijse}\ \emph {et~al.}(2008)\citenamefont {Huijse}, \citenamefont {Halverson}, \citenamefont {Fendley},\ and\ \citenamefont {Schoutens}}]{PhysRevLett.101.146406}%
  \BibitemOpen
  \bibfield  {author} {\bibinfo {author} {\bibfnamefont {L.}~\bibnamefont {Huijse}}, \bibinfo {author} {\bibfnamefont {J.}~\bibnamefont {Halverson}}, \bibinfo {author} {\bibfnamefont {P.}~\bibnamefont {Fendley}},\ and\ \bibinfo {author} {\bibfnamefont {K.}~\bibnamefont {Schoutens}},\ }\bibfield  {title} {\bibinfo {title} {Charge frustration and quantum criticality for strongly correlated fermions},\ }\href {https://doi.org/10.1103/PhysRevLett.101.146406} {\bibfield  {journal} {\bibinfo  {journal} {Phys. Rev. Lett.}\ }\textbf {\bibinfo {volume} {101}},\ \bibinfo {pages} {146406} (\bibinfo {year} {2008})}\BibitemShut {NoStop}%
\bibitem [{\citenamefont {Huijse}\ and\ \citenamefont {Schoutens}(2008)}]{huijse2008superfrustration}%
  \BibitemOpen
  \bibfield  {author} {\bibinfo {author} {\bibfnamefont {L.}~\bibnamefont {Huijse}}\ and\ \bibinfo {author} {\bibfnamefont {K.}~\bibnamefont {Schoutens}},\ }\bibfield  {title} {\bibinfo {title} {Superfrustration of charge degrees of freedom},\ }\href {https://link.springer.com/article/10.1140/epjb/e2008-00150-9} {\bibfield  {journal} {\bibinfo  {journal} {Euro. Phys. J. B}\ }\textbf {\bibinfo {volume} {64}},\ \bibinfo {pages} {543} (\bibinfo {year} {2008})}\BibitemShut {NoStop}%
\bibitem [{\citenamefont {Huijse}\ \emph {et~al.}(2012)\citenamefont {Huijse}, \citenamefont {Mehta}, \citenamefont {Moran}, \citenamefont {Schoutens},\ and\ \citenamefont {Vala}}]{huijse2012supersymmetric}%
  \BibitemOpen
  \bibfield  {author} {\bibinfo {author} {\bibfnamefont {L.}~\bibnamefont {Huijse}}, \bibinfo {author} {\bibfnamefont {D.}~\bibnamefont {Mehta}}, \bibinfo {author} {\bibfnamefont {N.}~\bibnamefont {Moran}}, \bibinfo {author} {\bibfnamefont {K.}~\bibnamefont {Schoutens}},\ and\ \bibinfo {author} {\bibfnamefont {J.}~\bibnamefont {Vala}},\ }\bibfield  {title} {\bibinfo {title} {Supersymmetric lattice fermions on the triangular lattice: superfrustration and criticality},\ }\href {https://iopscience.iop.org/article/10.1088/1367-2630/14/7/073002} {\bibfield  {journal} {\bibinfo  {journal} {New J. Phys.}\ }\textbf {\bibinfo {volume} {14}},\ \bibinfo {pages} {073002} (\bibinfo {year} {2012})}\BibitemShut {NoStop}%
\bibitem [{\citenamefont {Fendley}(2019)}]{fendley2019free}%
  \BibitemOpen
  \bibfield  {author} {\bibinfo {author} {\bibfnamefont {P.}~\bibnamefont {Fendley}},\ }\bibfield  {title} {\bibinfo {title} {Free fermions in disguise},\ }\href {https://iopscience.iop.org/article/10.1088/1751-8121/ab305d} {\bibfield  {journal} {\bibinfo  {journal} {J. Phys. A Math. Theor.}\ }\textbf {\bibinfo {volume} {52}},\ \bibinfo {pages} {335002} (\bibinfo {year} {2019})}\BibitemShut {NoStop}%
\bibitem [{\citenamefont {Chepiga}\ \emph {et~al.}(2021)\citenamefont {Chepiga}, \citenamefont {Min\'{a}\v{r}},\ and\ \citenamefont {Schoutens}}]{Chepiga-M-S-21}%
  \BibitemOpen
  \bibfield  {author} {\bibinfo {author} {\bibfnamefont {N.}~\bibnamefont {Chepiga}}, \bibinfo {author} {\bibfnamefont {J.}~\bibnamefont {Min\'{a}\v{r}}},\ and\ \bibinfo {author} {\bibfnamefont {K.}~\bibnamefont {Schoutens}},\ }\bibfield  {title} {\bibinfo {title} {{Supersymmetry and multicriticality in a ladder of constrained fermions}},\ }\href {https://doi.org/10.21468/SciPostPhys.11.3.059} {\bibfield  {journal} {\bibinfo  {journal} {SciPost Phys.}\ }\textbf {\bibinfo {volume} {11}},\ \bibinfo {pages} {059} (\bibinfo {year} {2021})}\BibitemShut {NoStop}%
\bibitem [{\citenamefont {Sannomiya}\ \emph {et~al.}(2016)\citenamefont {Sannomiya}, \citenamefont {Katsura},\ and\ \citenamefont {Nakayama}}]{sannomiya2016supersymmetry}%
  \BibitemOpen
  \bibfield  {author} {\bibinfo {author} {\bibfnamefont {N.}~\bibnamefont {Sannomiya}}, \bibinfo {author} {\bibfnamefont {H.}~\bibnamefont {Katsura}},\ and\ \bibinfo {author} {\bibfnamefont {Y.}~\bibnamefont {Nakayama}},\ }\bibfield  {title} {\bibinfo {title} {Supersymmetry breaking and {N}ambu-{G}oldstone fermions in an extended {N}icolai model},\ }\href {https://journals.aps.org/prd/abstract/10.1103/PhysRevD.94.045014} {\bibfield  {journal} {\bibinfo  {journal} {Phys. Rev. D}\ }\textbf {\bibinfo {volume} {94}},\ \bibinfo {pages} {045014} (\bibinfo {year} {2016})}\BibitemShut {NoStop}%
\bibitem [{\citenamefont {Sannomiya}\ \emph {et~al.}(2017)\citenamefont {Sannomiya}, \citenamefont {Katsura},\ and\ \citenamefont {Nakayama}}]{sannomiya2017supersymmetry}%
  \BibitemOpen
  \bibfield  {author} {\bibinfo {author} {\bibfnamefont {N.}~\bibnamefont {Sannomiya}}, \bibinfo {author} {\bibfnamefont {H.}~\bibnamefont {Katsura}},\ and\ \bibinfo {author} {\bibfnamefont {Y.}~\bibnamefont {Nakayama}},\ }\bibfield  {title} {\bibinfo {title} {Supersymmetry breaking and {N}ambu-{G}oldstone fermions with cubic dispersion},\ }\href {https://journals.aps.org/prd/abstract/10.1103/PhysRevD.95.065001} {\bibfield  {journal} {\bibinfo  {journal} {Phys. Rev. D}\ }\textbf {\bibinfo {volume} {95}},\ \bibinfo {pages} {065001} (\bibinfo {year} {2017})}\BibitemShut {NoStop}%
\bibitem [{\citenamefont {Moriya}(2018{\natexlab{a}})}]{moriya2018ergodicity}%
  \BibitemOpen
  \bibfield  {author} {\bibinfo {author} {\bibfnamefont {H.}~\bibnamefont {Moriya}},\ }\bibfield  {title} {\bibinfo {title} {Ergodicity breaking and localization of the {N}icolai supersymmetric fermion lattice model},\ }\href {https://link.springer.com/article/10.1007/s10955-018-2100-3} {\bibfield  {journal} {\bibinfo  {journal} {J. Stat. Phys.}\ }\textbf {\bibinfo {volume} {172}},\ \bibinfo {pages} {1270} (\bibinfo {year} {2018}{\natexlab{a}})}\BibitemShut {NoStop}%
\bibitem [{\citenamefont {Moriya}(2018{\natexlab{b}})}]{moriya2018supersymmetry}%
  \BibitemOpen
  \bibfield  {author} {\bibinfo {author} {\bibfnamefont {H.}~\bibnamefont {Moriya}},\ }\bibfield  {title} {\bibinfo {title} {Supersymmetry breakdown for an extended version of the {N}icolai supersymmetric fermion lattice model},\ }\href {https://journals.aps.org/prd/abstract/10.1103/PhysRevD.98.015018} {\bibfield  {journal} {\bibinfo  {journal} {Phys. Rev. D}\ }\textbf {\bibinfo {volume} {98}},\ \bibinfo {pages} {015018} (\bibinfo {year} {2018}{\natexlab{b}})}\BibitemShut {NoStop}%
\bibitem [{\citenamefont {Sannomiya}\ and\ \citenamefont {Katsura}(2019)}]{sannomiya2019supersymmetry}%
  \BibitemOpen
  \bibfield  {author} {\bibinfo {author} {\bibfnamefont {N.}~\bibnamefont {Sannomiya}}\ and\ \bibinfo {author} {\bibfnamefont {H.}~\bibnamefont {Katsura}},\ }\bibfield  {title} {\bibinfo {title} {Supersymmetry breaking and {N}ambu-{G}oldstone fermions in interacting {M}ajorana chains},\ }\href {https://journals.aps.org/prd/abstract/10.1103/PhysRevD.99.045002} {\bibfield  {journal} {\bibinfo  {journal} {Phys. Rev. D}\ }\textbf {\bibinfo {volume} {99}},\ \bibinfo {pages} {045002} (\bibinfo {year} {2019})}\BibitemShut {NoStop}%
\bibitem [{\citenamefont {Sannomiya}(2021)}]{sannomiyaDron}%
  \BibitemOpen
  \bibfield  {author} {\bibinfo {author} {\bibfnamefont {N.}~\bibnamefont {Sannomiya}},\ }\emph {\bibinfo {title} {Spontaneous Supersymmetry Breaking and {N}ambu-{G}oldstone Modes in Interacting {M}ajorana Chains}},\ \href {https://park.itc.u-tokyo.ac.jp/hkatsura-lab/thesis.html} {Ph.D. thesis},\ \bibinfo  {school} {The University of Tokyo} (\bibinfo {year} {2021})\BibitemShut {NoStop}%
\bibitem [{\citenamefont {Katsura}\ \emph {et~al.}(2020)\citenamefont {Katsura}, \citenamefont {Moriya},\ and\ \citenamefont {Nakayama}}]{Katsura_2020}%
  \BibitemOpen
  \bibfield  {author} {\bibinfo {author} {\bibfnamefont {H.}~\bibnamefont {Katsura}}, \bibinfo {author} {\bibfnamefont {H.}~\bibnamefont {Moriya}},\ and\ \bibinfo {author} {\bibfnamefont {Y.}~\bibnamefont {Nakayama}},\ }\bibfield  {title} {\bibinfo {title} {Characterization of degenerate supersymmetric ground states of the {N}icolai supersymmetric fermion lattice model by symmetry breakdown},\ }\href {https://doi.org/10.1088/1751-8121/ab9916} {\bibfield  {journal} {\bibinfo  {journal} {J. Phys. A: Math. Theor.}\ }\textbf {\bibinfo {volume} {53}},\ \bibinfo {pages} {385003} (\bibinfo {year} {2020})}\BibitemShut {NoStop}%
\bibitem [{\citenamefont {Katsura}\ and\ \citenamefont {Nakayama}(2022)}]{SUSYfracton}%
  \BibitemOpen
  \bibfield  {author} {\bibinfo {author} {\bibfnamefont {H.}~\bibnamefont {Katsura}}\ and\ \bibinfo {author} {\bibfnamefont {Y.}~\bibnamefont {Nakayama}},\ }\bibfield  {title} {\bibinfo {title} {Spontaneously broken supersymmetric fracton phases with fermionic subsystem symmetries},\ }\href {https://link.springer.com/article/10.1007/JHEP08(2022)072} {J. High Energy Phys. \textbf{08} (2022) 072}\BibitemShut {NoStop}%
\bibitem [{\citenamefont {Miura}\ and\ \citenamefont {Totsuka}(2023)}]{miura2023supersymmetry}%
  \BibitemOpen
  \bibfield  {author} {\bibinfo {author} {\bibfnamefont {U.}~\bibnamefont {Miura}}\ and\ \bibinfo {author} {\bibfnamefont {K.}~\bibnamefont {Totsuka}},\ }\bibfield  {title} {\bibinfo {title} {Supersymmetry breaking in a generalized {N}icolai model with fermion pairing},\ }\href {https://arxiv.org/abs/2308.03346} {\bibinfo {journal} {arXiv:2308.03346}}\BibitemShut {NoStop}%
\bibitem [{\citenamefont {Miura}\ \emph {et~al.}(2024)\citenamefont {Miura}, \citenamefont {Shimomura},\ and\ \citenamefont {Totsuka}}]{miura2024interacting}%
  \BibitemOpen
  \bibfield  {author} {\bibinfo {author} {\bibfnamefont {U.}~\bibnamefont {Miura}}, \bibinfo {author} {\bibfnamefont {K.}~\bibnamefont {Shimomura}},\ and\ \bibinfo {author} {\bibfnamefont {K.}~\bibnamefont {Totsuka}},\ }\bibfield  {title} {\bibinfo {title} {Interacting {K}itaev chain with $\mathcal{N}=1$ supersymmetry},\ }\href {https://journals.aps.org/prb/abstract/10.1103/PhysRevB.109.085141} {\bibfield  {journal} {\bibinfo  {journal} {Phys. Rev. B}\ }\textbf {\bibinfo {volume} {109}},\ \bibinfo {pages} {085141} (\bibinfo {year} {2024})}\BibitemShut {NoStop}%
\bibitem [{\citenamefont {Grover}\ \emph {et~al.}(2014)\citenamefont {Grover}, \citenamefont {Sheng},\ and\ \citenamefont {Vishwanath}}]{grover2014emergent}%
  \BibitemOpen
  \bibfield  {author} {\bibinfo {author} {\bibfnamefont {T.}~\bibnamefont {Grover}}, \bibinfo {author} {\bibfnamefont {D.}~\bibnamefont {Sheng}},\ and\ \bibinfo {author} {\bibfnamefont {A.}~\bibnamefont {Vishwanath}},\ }\bibfield  {title} {\bibinfo {title} {Emergent space-time supersymmetry at the boundary of a topological phase},\ }\href {https://www.science.org/doi/abs/10.1126/science.1248253} {\bibfield  {journal} {\bibinfo  {journal} {Science}\ }\textbf {\bibinfo {volume} {344}},\ \bibinfo {pages} {280} (\bibinfo {year} {2014})}\BibitemShut {NoStop}%
\bibitem [{\citenamefont {Hsieh}\ \emph {et~al.}(2016)\citenamefont {Hsieh}, \citenamefont {Hal\'{a}sz},\ and\ \citenamefont {Grover}}]{hsieh2016all}%
  \BibitemOpen
  \bibfield  {author} {\bibinfo {author} {\bibfnamefont {T.~H.}\ \bibnamefont {Hsieh}}, \bibinfo {author} {\bibfnamefont {G.~B.}\ \bibnamefont {Hal\'{a}sz}},\ and\ \bibinfo {author} {\bibfnamefont {T.}~\bibnamefont {Grover}},\ }\bibfield  {title} {\bibinfo {title} {All {M}ajorana models with translation symmetry are supersymmetric},\ }\href {https://doi.org/10.1103/PhysRevLett.117.166802} {\bibfield  {journal} {\bibinfo  {journal} {Phys. Rev. Lett.}\ }\textbf {\bibinfo {volume} {117}},\ \bibinfo {pages} {166802} (\bibinfo {year} {2016})}\BibitemShut {NoStop}%
\bibitem [{\citenamefont {Marra}\ \emph {et~al.}(2022{\natexlab{a}})\citenamefont {Marra}, \citenamefont {Inotani},\ and\ \citenamefont {Nitta}}]{marra20221d}%
  \BibitemOpen
  \bibfield  {author} {\bibinfo {author} {\bibfnamefont {P.}~\bibnamefont {Marra}}, \bibinfo {author} {\bibfnamefont {D.}~\bibnamefont {Inotani}},\ and\ \bibinfo {author} {\bibfnamefont {M.}~\bibnamefont {Nitta}},\ }\bibfield  {title} {\bibinfo {title} {1{D} {M}ajorana {G}oldstinos and partial supersymmetry breaking in quantum wires},\ }\href {https://www.nature.com/articles/s42005-022-00920-4} {\bibfield  {journal} {\bibinfo  {journal} {Commun. Phys.}\ }\textbf {\bibinfo {volume} {5}},\ \bibinfo {pages} {149} (\bibinfo {year} {2022}{\natexlab{a}})}\BibitemShut {NoStop}%
\bibitem [{\citenamefont {Marra}\ \emph {et~al.}(2022{\natexlab{b}})\citenamefont {Marra}, \citenamefont {Inotani},\ and\ \citenamefont {Nitta}}]{PhysRevB.105.214525}%
  \BibitemOpen
  \bibfield  {author} {\bibinfo {author} {\bibfnamefont {P.}~\bibnamefont {Marra}}, \bibinfo {author} {\bibfnamefont {D.}~\bibnamefont {Inotani}},\ and\ \bibinfo {author} {\bibfnamefont {M.}~\bibnamefont {Nitta}},\ }\bibfield  {title} {\bibinfo {title} {Dispersive one-dimensional majorana modes with emergent supersymmetry in one-dimensional proximitized superconductors via spatially modulated potentials and magnetic fields},\ }\href {https://doi.org/10.1103/PhysRevB.105.214525} {\bibfield  {journal} {\bibinfo  {journal} {Phys. Rev. B}\ }\textbf {\bibinfo {volume} {105}},\ \bibinfo {pages} {214525} (\bibinfo {year} {2022}{\natexlab{b}})}\BibitemShut {NoStop}%
\bibitem [{\citenamefont {Marra}\ \emph {et~al.}(2024)\citenamefont {Marra}, \citenamefont {Inotani}, \citenamefont {Mizushima},\ and\ \citenamefont {Nitta}}]{marra2024majorana}%
  \BibitemOpen
  \bibfield  {author} {\bibinfo {author} {\bibfnamefont {P.}~\bibnamefont {Marra}}, \bibinfo {author} {\bibfnamefont {D.}~\bibnamefont {Inotani}}, \bibinfo {author} {\bibfnamefont {T.}~\bibnamefont {Mizushima}},\ and\ \bibinfo {author} {\bibfnamefont {M.}~\bibnamefont {Nitta}},\ }\bibfield  {title} {\bibinfo {title} {{M}ajorana modes in striped two-dimensional inhomogeneous topological superconductors},\ }\href {https://www.nature.com/articles/s41535-024-00672-0} {\bibfield  {journal} {\bibinfo  {journal} {npj Quantum Mater.}\ }\textbf {\bibinfo {volume} {9}},\ \bibinfo {pages} {59} (\bibinfo {year} {2024})}\BibitemShut {NoStop}%
\bibitem [{\citenamefont {Hidaka}(2013)}]{PhysRevLett.110.091601}%
  \BibitemOpen
  \bibfield  {author} {\bibinfo {author} {\bibfnamefont {Y.}~\bibnamefont {Hidaka}},\ }\bibfield  {title} {\bibinfo {title} {Counting rule for {N}ambu-{G}oldstone modes in nonrelativistic systems},\ }\href {https://doi.org/10.1103/PhysRevLett.110.091601} {\bibfield  {journal} {\bibinfo  {journal} {Phys. Rev. Lett.}\ }\textbf {\bibinfo {volume} {110}},\ \bibinfo {pages} {091601} (\bibinfo {year} {2013})}\BibitemShut {NoStop}%
\bibitem [{\citenamefont {Watanabe}\ and\ \citenamefont {Murayama}(2012)}]{PhysRevLett.108.251602}%
  \BibitemOpen
  \bibfield  {author} {\bibinfo {author} {\bibfnamefont {H.}~\bibnamefont {Watanabe}}\ and\ \bibinfo {author} {\bibfnamefont {H.}~\bibnamefont {Murayama}},\ }\bibfield  {title} {\bibinfo {title} {Unified description of {N}ambu-{G}oldstone bosons without lorentz invariance},\ }\href {https://doi.org/10.1103/PhysRevLett.108.251602} {\bibfield  {journal} {\bibinfo  {journal} {Phys. Rev. Lett.}\ }\textbf {\bibinfo {volume} {108}},\ \bibinfo {pages} {251602} (\bibinfo {year} {2012})}\BibitemShut {NoStop}%
\bibitem [{\citenamefont {Rahmani}\ \emph {et~al.}(2015{\natexlab{a}})\citenamefont {Rahmani}, \citenamefont {Zhu}, \citenamefont {Franz},\ and\ \citenamefont {Affleck}}]{rahmani2015emergent}%
  \BibitemOpen
  \bibfield  {author} {\bibinfo {author} {\bibfnamefont {A.}~\bibnamefont {Rahmani}}, \bibinfo {author} {\bibfnamefont {X.}~\bibnamefont {Zhu}}, \bibinfo {author} {\bibfnamefont {M.}~\bibnamefont {Franz}},\ and\ \bibinfo {author} {\bibfnamefont {I.}~\bibnamefont {Affleck}},\ }\bibfield  {title} {\bibinfo {title} {Emergent supersymmetry from strongly interacting {M}ajorana zero modes},\ }\href {https://journals.aps.org/prl/abstract/10.1103/PhysRevLett.115.166401} {\bibfield  {journal} {\bibinfo  {journal} {Phys. Rev. Lett.}\ }\textbf {\bibinfo {volume} {115}},\ \bibinfo {pages} {166401} (\bibinfo {year} {2015}{\natexlab{a}})}\BibitemShut {NoStop}%
\bibitem [{\citenamefont {Rahmani}\ and\ \citenamefont {Franz}(2019)}]{rahmani2019interacting}%
  \BibitemOpen
  \bibfield  {author} {\bibinfo {author} {\bibfnamefont {A.}~\bibnamefont {Rahmani}}\ and\ \bibinfo {author} {\bibfnamefont {M.}~\bibnamefont {Franz}},\ }\bibfield  {title} {\bibinfo {title} {Interacting {M}ajorana fermions},\ }\href {https://iopscience.iop.org/article/10.1088/1361-6633/ab28ef/meta} {\bibfield  {journal} {\bibinfo  {journal} {Reports on Progress in Physics}\ }\textbf {\bibinfo {volume} {82}},\ \bibinfo {pages} {084501} (\bibinfo {year} {2019})}\BibitemShut {NoStop}%
\bibitem [{\citenamefont {Rahmani}\ \emph {et~al.}(2015{\natexlab{b}})\citenamefont {Rahmani}, \citenamefont {Zhu}, \citenamefont {Franz},\ and\ \citenamefont {Affleck}}]{PhysRevB.92.235123}%
  \BibitemOpen
  \bibfield  {author} {\bibinfo {author} {\bibfnamefont {A.}~\bibnamefont {Rahmani}}, \bibinfo {author} {\bibfnamefont {X.}~\bibnamefont {Zhu}}, \bibinfo {author} {\bibfnamefont {M.}~\bibnamefont {Franz}},\ and\ \bibinfo {author} {\bibfnamefont {I.}~\bibnamefont {Affleck}},\ }\bibfield  {title} {\bibinfo {title} {Phase diagram of the interacting {M}ajorana chain model},\ }\href {https://doi.org/10.1103/PhysRevB.92.235123} {\bibfield  {journal} {\bibinfo  {journal} {Phys. Rev. B}\ }\textbf {\bibinfo {volume} {92}},\ \bibinfo {pages} {235123} (\bibinfo {year} {2015}{\natexlab{b}})}\BibitemShut {NoStop}%
\bibitem [{\citenamefont {O'Brien}\ and\ \citenamefont {Fendley}(2018)}]{o2018lattice}%
  \BibitemOpen
  \bibfield  {author} {\bibinfo {author} {\bibfnamefont {E.}~\bibnamefont {O'Brien}}\ and\ \bibinfo {author} {\bibfnamefont {P.}~\bibnamefont {Fendley}},\ }\bibfield  {title} {\bibinfo {title} {Lattice supersymmetry and order-disorder coexistence in the tricritical {I}sing model},\ }\href {https://doi.org/10.1103/PhysRevLett.120.206403} {\bibfield  {journal} {\bibinfo  {journal} {Phys. Rev. Lett.}\ }\textbf {\bibinfo {volume} {120}},\ \bibinfo {pages} {206403} (\bibinfo {year} {2018})}\BibitemShut {NoStop}%
\bibitem [{\citenamefont {Kitaev}(2001)}]{kitaev2001unpaired}%
  \BibitemOpen
  \bibfield  {author} {\bibinfo {author} {\bibfnamefont {A.~Y.}\ \bibnamefont {Kitaev}},\ }\bibfield  {title} {\bibinfo {title} {Unpaired {M}ajorana fermions in quantum wires},\ }\href {https://iopscience.iop.org/article/10.1070/1063-7869/44/10S/S29} {\bibfield  {journal} {\bibinfo  {journal} {Soviet {P}hysics. Uspekhi}\ }\textbf {\bibinfo {volume} {44}},\ \bibinfo {pages} {131} (\bibinfo {year} {2001})}\BibitemShut {NoStop}%
\bibitem [{\citenamefont {Zamolodchikov}(1987)}]{Zamolodchikov-c-87}%
  \BibitemOpen
  \bibfield  {author} {\bibinfo {author} {\bibfnamefont {A.~B.}\ \bibnamefont {Zamolodchikov}},\ }\bibfield  {title} {\bibinfo {title} {Renormalization group and perturbation theory about fixed points in two-dimensional field theory},\ }\href {http://inis.iaea.org/search/search.aspx?orig_q=RN:19090153} {\bibfield  {journal} {\bibinfo  {journal} {Sov. J. Nucl. Phys.}\ }\textbf {\bibinfo {volume} {46}},\ \bibinfo {pages} {1090} (\bibinfo {year} {1987})}\BibitemShut {NoStop}%
\bibitem [{\citenamefont {Kastor}\ \emph {et~al.}(1989)\citenamefont {Kastor}, \citenamefont {Martinec},\ and\ \citenamefont {Shenker}}]{Kastor-M-S-89}%
  \BibitemOpen
  \bibfield  {author} {\bibinfo {author} {\bibfnamefont {D.~A.}\ \bibnamefont {Kastor}}, \bibinfo {author} {\bibfnamefont {E.~J.}\ \bibnamefont {Martinec}},\ and\ \bibinfo {author} {\bibfnamefont {S.~H.}\ \bibnamefont {Shenker}},\ }\bibfield  {title} {\bibinfo {title} {{RG} flow in {$N = 1$} discrete series},\ }\href {https://doi.org/10.1016/0550-3213(89)90060-6} {\bibfield  {journal} {\bibinfo  {journal} {Nucl. Phys. B}\ }\textbf {\bibinfo {volume} {316}},\ \bibinfo {pages} {590} (\bibinfo {year} {1989})}\BibitemShut {NoStop}%
\bibitem [{\citenamefont {Zamolodchikov}(1991)}]{Zamolodchikov-TCIM-Ising-91}%
  \BibitemOpen
  \bibfield  {author} {\bibinfo {author} {\bibfnamefont {A.~B.}\ \bibnamefont {Zamolodchikov}},\ }\bibfield  {title} {\bibinfo {title} {From tricritical {I}sing to critical {I}sing by thermodynamic {B}ethe ansatz},\ }\href {https://doi.org/10.1016/0550-3213(91)90423-U} {\bibfield  {journal} {\bibinfo  {journal} {Nucl. Phys. B}\ }\textbf {\bibinfo {volume} {358}},\ \bibinfo {pages} {524} (\bibinfo {year} {1991})}\BibitemShut {NoStop}%
\bibitem [{\citenamefont {Sannomiya}(2024)}]{sannomiya2024spontaneous}%
  \BibitemOpen
  \bibfield  {author} {\bibinfo {author} {\bibfnamefont {N.}~\bibnamefont {Sannomiya}},\ }\bibfield  {title} {\bibinfo {title} {Spontaneous supersymmetry breaking and {N}ambu-{G}oldstone modes in interacting {M}ajorana chains},\ }\href {https://arxiv.org/abs/2401.07419} {\bibinfo {journal} {arXiv:2401.07419}}\BibitemShut {NoStop}%
\bibitem [{\citenamefont {Salam}\ and\ \citenamefont {Strathdee}(1974)}]{Salam-S-74}%
  \BibitemOpen
  \bibfield  {author} {\bibinfo {author} {\bibfnamefont {A.}~\bibnamefont {Salam}}\ and\ \bibinfo {author} {\bibfnamefont {J.}~\bibnamefont {Strathdee}},\ }\bibfield  {title} {\bibinfo {title} {On goldstone fermions},\ }\href {https://doi.org/10.1016/0370-2693(74)90637-6} {\bibfield  {journal} {\bibinfo  {journal} {Phys. Lett. B}\ }\textbf {\bibinfo {volume} {49}},\ \bibinfo {pages} {465} (\bibinfo {year} {1974})}\BibitemShut {NoStop}%
\bibitem [{\citenamefont {White}(1992)}]{PhysRevLett.69.2863}%
  \BibitemOpen
  \bibfield  {author} {\bibinfo {author} {\bibfnamefont {S.~R.}\ \bibnamefont {White}},\ }\bibfield  {title} {\bibinfo {title} {Density matrix formulation for quantum renormalization groups},\ }\href {https://doi.org/10.1103/PhysRevLett.69.2863} {\bibfield  {journal} {\bibinfo  {journal} {Phys. Rev. Lett.}\ }\textbf {\bibinfo {volume} {69}},\ \bibinfo {pages} {2863} (\bibinfo {year} {1992})}\BibitemShut {NoStop}%
\bibitem [{\citenamefont {Fishman}\ \emph {et~al.}(2022)\citenamefont {Fishman}, \citenamefont {White},\ and\ \citenamefont {Stoudenmire}}]{iTensor}%
  \BibitemOpen
  \bibfield  {author} {\bibinfo {author} {\bibfnamefont {M.}~\bibnamefont {Fishman}}, \bibinfo {author} {\bibfnamefont {S.~R.}\ \bibnamefont {White}},\ and\ \bibinfo {author} {\bibfnamefont {E.~M.}\ \bibnamefont {Stoudenmire}},\ }\bibfield  {title} {\bibinfo {title} {{Codebase release 0.3 for ITensor}},\ }\href {https://doi.org/10.21468/SciPostPhysCodeb.4-r0.3} {\bibfield  {journal} {\bibinfo  {journal} {SciPost Phys. Codebases}\ ,\ \bibinfo {pages} {4}} (\bibinfo {year} {2022})}\BibitemShut {NoStop}%
\bibitem [{Note1()}]{note-gap-definition}%
  \BibitemOpen
\bibinfo {note} {Because of the even degeneracy of the positive energy states caused by supersymmetry, the ground state of finite systems is also degenerate. Specifically, in the region $0<\alpha<1/2$, the ground state is doubly degenerate. Therefore, $\Delta(L;\alpha)$ is defined as described in the main text, and was calculated numerically.}\BibitemShut {Stop}%
\bibitem [{\citenamefont {Blume}\ \emph {et~al.}(1971)\citenamefont {Blume}, \citenamefont {Emery},\ and\ \citenamefont {Griffiths}}]{PhysRevA.4.1071}%
  \BibitemOpen
  \bibfield  {author} {\bibinfo {author} {\bibfnamefont {M.}~\bibnamefont {Blume}}, \bibinfo {author} {\bibfnamefont {V.~J.}\ \bibnamefont {Emery}},\ and\ \bibinfo {author} {\bibfnamefont {R.~B.}\ \bibnamefont {Griffiths}},\ }\bibfield  {title} {\bibinfo {title} {{I}sing model for the lambda transition and phase separation in {He${}^{3}$-He${}^{4}$} mixtures},\ }\href {https://doi.org/10.1103/PhysRevA.4.1071} {\bibfield  {journal} {\bibinfo  {journal} {Phys. Rev. A}\ }\textbf {\bibinfo {volume} {4}},\ \bibinfo {pages} {1071} (\bibinfo {year} {1971})}\BibitemShut {NoStop}%
\bibitem [{\citenamefont {Ludwig}\ and\ \citenamefont {Cardy}(1987)}]{Ludwig-C-87}%
  \BibitemOpen
  \bibfield  {author} {\bibinfo {author} {\bibfnamefont {A.~W.}\ \bibnamefont {Ludwig}}\ and\ \bibinfo {author} {\bibfnamefont {J.~L.}\ \bibnamefont {Cardy}},\ }\bibfield  {title} {\bibinfo {title} {Perturbative evaluation of the conformal anomaly at new critical points with applications to random systems},\ }\href {https://doi.org/http://dx.doi.org/10.1016/0550-3213(87)90362-2} {\bibfield  {journal} {\bibinfo  {journal} {Nucl. Phys. B}\ }\textbf {\bibinfo {volume} {285}},\ \bibinfo {pages} {687 } (\bibinfo {year} {1987})}\BibitemShut {NoStop}%
\bibitem [{\citenamefont {Zamolodchikov}(1991)}]{Zamolodchikov-RSOS-TBA-91}%
  \BibitemOpen
  \bibfield {author} {\bibinfo {author} {\bibfnamefont {A.~B.}\ \bibnamefont {Zamolodchikov}},\ }\bibfield {title} {\bibinfo {title} {Thermodynamic {B}ethe ansatz for {RSOS} scattering theories},\ }\href {https://doi.org/10.1016/0550-3213(91)90422-D} {\bibfield {journal} {\bibinfo {journal} {Nucl. Phys. B}\ }\textbf {\bibinfo {volume} {358}},\ \bibinfo {pages} {497} (\bibinfo {year} {1991})}\BibitemShut {NoStop}%
\bibitem [{\citenamefont {Lepori}\ \emph {et~al.}(2008)\citenamefont {Lepori}, \citenamefont {Mussardo},\ and\ \citenamefont {Tóth}}]{Lepori_2008}%
  \BibitemOpen
  \bibfield  {author} {\bibinfo {author} {\bibfnamefont {L.}~\bibnamefont {Lepori}}, \bibinfo {author} {\bibfnamefont {G.}~\bibnamefont {Mussardo}},\ and\ \bibinfo {author} {\bibfnamefont {G.~Z.}\ \bibnamefont {Tóth}},\ }\bibfield  {title} {\bibinfo {title} {The particle spectrum of the tricritical {I}sing model with spin reversal symmetric perturbations},\ }\href {https://doi.org/10.1088/1742-5468/2008/09/P09004} {\bibfield  {journal} {\bibinfo  {journal} {J. Stat. Mech. Theory Exp.}\ }\textbf {\bibinfo {volume} {2008}},\ \bibinfo {pages} {P09004} (\bibinfo {year} {2008})}\BibitemShut {NoStop}%
\bibitem [{\citenamefont {L\"{a}ssig}\ \emph {et~al.}(1991)\citenamefont {L\"{a}ssig}, \citenamefont {Mussardo},\ and\ \citenamefont {Cardy}}]{Laessig-M-C-91}%
  \BibitemOpen
  \bibfield  {author} {\bibinfo {author} {\bibfnamefont {M.}~\bibnamefont {L\"{a}ssig}}, \bibinfo {author} {\bibfnamefont {G.}~\bibnamefont {Mussardo}},\ and\ \bibinfo {author} {\bibfnamefont {J.~L.}\ \bibnamefont {Cardy}},\ }\bibfield  {title} {\bibinfo {title} {The scaling region of the tricritical {I}sing model in two dimensions},\ }\href {https://doi.org/10.1016/0550-3213(91)90206-D} {\bibfield  {journal} {\bibinfo  {journal} {Nucl. Phys. B}\ }\textbf {\bibinfo {volume} {348}},\ \bibinfo {pages} {591} (\bibinfo {year} {1991})}\BibitemShut {NoStop}%
\bibitem [{\citenamefont {Holzhey}\ \emph {et~al.}(1994)\citenamefont {Holzhey}, \citenamefont {Larsen},\ and\ \citenamefont {Wilczek}}]{HOLZHEY1994443}%
  \BibitemOpen
  \bibfield  {author} {\bibinfo {author} {\bibfnamefont {C.}~\bibnamefont {Holzhey}}, \bibinfo {author} {\bibfnamefont {F.}~\bibnamefont {Larsen}},\ and\ \bibinfo {author} {\bibfnamefont {F.}~\bibnamefont {Wilczek}},\ }\bibfield  {title} {\bibinfo {title} {Geometric and renormalized entropy in conformal field theory},\ }\href {https://doi.org/https://doi.org/10.1016/0550-3213(94)90402-2} {\bibfield  {journal} {\bibinfo  {journal} {Nuclear Physics B}\ }\textbf {\bibinfo {volume} {424}},\ \bibinfo {pages} {443} (\bibinfo {year} {1994})}\BibitemShut {NoStop}%
\bibitem [{\citenamefont {Vidal}\ \emph {et~al.}(2003)\citenamefont {Vidal}, \citenamefont {Latorre}, \citenamefont {Rico},\ and\ \citenamefont {Kitaev}}]{PhysRevLett.90.227902}%
  \BibitemOpen
  \bibfield  {author} {\bibinfo {author} {\bibfnamefont {G.}~\bibnamefont {Vidal}}, \bibinfo {author} {\bibfnamefont {J.~I.}\ \bibnamefont {Latorre}}, \bibinfo {author} {\bibfnamefont {E.}~\bibnamefont {Rico}},\ and\ \bibinfo {author} {\bibfnamefont {A.}~\bibnamefont {Kitaev}},\ }\bibfield  {title} {\bibinfo {title} {Entanglement in quantum critical phenomena},\ }\href {https://doi.org/10.1103/PhysRevLett.90.227902} {\bibfield  {journal} {\bibinfo  {journal} {Phys. Rev. Lett.}\ }\textbf {\bibinfo {volume} {90}},\ \bibinfo {pages} {227902} (\bibinfo {year} {2003})}\BibitemShut {NoStop}%
\bibitem [{\citenamefont {Calabrese}\ and\ \citenamefont {Cardy}(2004)}]{PasqualeCalabrese_2004}%
  \BibitemOpen
  \bibfield  {author} {\bibinfo {author} {\bibfnamefont {P.}~\bibnamefont {Calabrese}}\ and\ \bibinfo {author} {\bibfnamefont {J.}~\bibnamefont {Cardy}},\ }\bibfield  {title} {\bibinfo {title} {Entanglement entropy and quantum field theory},\ }\href {https://doi.org/10.1088/1742-5468/2004/06/P06002} {\bibfield  {journal} {\bibinfo  {journal} {J. Stat. Mech. Theory Exp.}\ }\textbf {\bibinfo {volume} {2004}},\ \bibinfo {pages} {P06002} (\bibinfo {year} {2004})}\BibitemShut {NoStop}%
\bibitem [{\citenamefont {Xavier}\ and\ \citenamefont {Alcaraz}(2011)}]{PhysRevB.84.094410}%
  \BibitemOpen
  \bibfield  {author} {\bibinfo {author} {\bibfnamefont {J.~C.}\ \bibnamefont {Xavier}}\ and\ \bibinfo {author} {\bibfnamefont {F.~C.}\ \bibnamefont {Alcaraz}},\ }\bibfield  {title} {\bibinfo {title} {Precise determination of quantum critical points by the violation of the entropic area law},\ }\href {https://doi.org/10.1103/PhysRevB.84.094410} {\bibfield  {journal} {\bibinfo  {journal} {Phys. Rev. B}\ }\textbf {\bibinfo {volume} {84}},\ \bibinfo {pages} {094410} (\bibinfo {year} {2011})}\BibitemShut {NoStop}%
\bibitem [{\citenamefont {Cardy}(1986{\natexlab{a}})}]{Cardy-86}%
  \BibitemOpen
  \bibfield  {author} {\bibinfo {author} {\bibfnamefont {J.~L.}\ \bibnamefont {Cardy}},\ }\bibfield  {title} {\bibinfo {title} {Operator content of two-dimensional conformally invariant theories},\ }\href {https://doi.org/http://dx.doi.org/10.1016/0550-3213(86)90552-3} {\bibfield  {journal} {\bibinfo  {journal} {Nucl. Phys. B}\ }\textbf {\bibinfo {volume} {270}},\ \bibinfo {pages} {186 } (\bibinfo {year} {1986}{\natexlab{a}})}\BibitemShut {NoStop}%
\bibitem [{\citenamefont {Cardy}(1986{\natexlab{b}})}]{Cardy-86-bc}%
  \BibitemOpen
  \bibfield  {author} {\bibinfo {author} {\bibfnamefont {J.~L.}\ \bibnamefont {Cardy}},\ }\bibfield  {title} {\bibinfo {title} {Effect of boundary conditions on the operator content of two-dimensional conformally invariant theories},\ }\href {https://doi.org/http://dx.doi.org/10.1016/0550-3213(86)90596-1} {\bibfield  {journal} {\bibinfo  {journal} {Nucl. Phys. B}\ }\textbf {\bibinfo {volume} {275}},\ \bibinfo {pages} {200 } (\bibinfo {year} {1986}{\natexlab{b}})}\BibitemShut {NoStop}%
\bibitem [{\citenamefont {Di~Francesco}\ \emph {et~al.}(1996)\citenamefont {Di~Francesco}, \citenamefont {Mathieu},\ and\ \citenamefont {S\'{e}n\'{e}chal}}]{DiFrancesco-M-S-book}%
  \BibitemOpen
  \bibfield  {author} {\bibinfo {author} {\bibfnamefont {P.}~\bibnamefont {Di~Francesco}}, \bibinfo {author} {\bibfnamefont {P.}~\bibnamefont {Mathieu}},\ and\ \bibinfo {author} {\bibfnamefont {D.}~\bibnamefont {S\'{e}n\'{e}chal}},\ }\href@noop {} {\emph {\bibinfo {title} {Conformal Field Theory}}}\ (\bibinfo  {publisher} {Springer Verlag},\ \bibinfo {address} {Berlin},\ \bibinfo {year} {1996})\BibitemShut {NoStop}%
\bibitem [{Note2()}]{note-momentum-resolved}%
  \BibitemOpen
  \bibinfo {note} {To obtain the momentum-resolved spectrum, we need to keep $x$ finite.}\BibitemShut {Stop}%
\bibitem [{\citenamefont {Cappelli}\ \emph {et~al.}(1987{\natexlab{a}})\citenamefont {Cappelli}, \citenamefont {Itzykson},\ and\ \citenamefont {Zuber}}]{Cappelli-I-Z-NP-87}%
  \BibitemOpen
  \bibfield  {author} {\bibinfo {author} {\bibfnamefont {A.}~\bibnamefont {Cappelli}}, \bibinfo {author} {\bibfnamefont {C.}~\bibnamefont {Itzykson}},\ and\ \bibinfo {author} {\bibfnamefont {J.-B.}\ \bibnamefont {Zuber}},\ }\bibfield  {title} {\bibinfo {title} {Modular invariant partition functions in two dimensions},\ }\href {https://doi.org/10.1016/0550-3213(87)90155-6} {\bibfield  {journal} {\bibinfo  {journal} {Nucl. Phys. B}\ }\textbf {\bibinfo {volume} {280}},\ \bibinfo {pages} {445} (\bibinfo {year} {1987}{\natexlab{a}})}\BibitemShut {NoStop}%
\bibitem [{\citenamefont {Cappelli}\ \emph {et~al.}(1987{\natexlab{b}})\citenamefont {Cappelli}, \citenamefont {Itzykson},\ and\ \citenamefont {Zuber}}]{Cappelli-I-Z-CMP-87}%
  \BibitemOpen
  \bibfield  {author} {\bibinfo {author} {\bibfnamefont {A.}~\bibnamefont {Cappelli}}, \bibinfo {author} {\bibfnamefont {C.}~\bibnamefont {Itzykson}},\ and\ \bibinfo {author} {\bibfnamefont {J.~B.}\ \bibnamefont {Zuber}},\ }\bibfield  {title} {\bibinfo {title} {The {A-D-E} classification of minimal and {$A_{1}^{(1)}$} conformal invariant theories},\ }\href {https://doi.org/10.1007/BF01221394} {\bibfield  {journal} {\bibinfo  {journal} {Comm. Math. Phys.}\ }\textbf {\bibinfo {volume} {113}},\ \bibinfo {pages} {1} (\bibinfo {year} {1987}{\natexlab{b}})}\BibitemShut {NoStop}%
\bibitem [{\citenamefont {Cappelli}(1987)}]{CAPPELLI198782}%
  \BibitemOpen
  \bibfield  {author} {\bibinfo {author} {\bibfnamefont {A.}~\bibnamefont {Cappelli}},\ }\bibfield  {title} {\bibinfo {title} {Modular invariant partition functions of superconformal theories},\ }\href {https://doi.org/https://doi.org/10.1016/0370-2693(87)91532-2} {\bibfield  {journal} {\bibinfo  {journal} {Phys. Lett. B}\ }\textbf {\bibinfo {volume} {185}},\ \bibinfo {pages} {82} (\bibinfo {year} {1987})}\BibitemShut {NoStop}%
\end{thebibliography}
\end{document}